%% file: arp82dauaibec-cleaned.tex
\RequirePackage[l2tabu, orthodox]{nag}
\RequirePackage{snapshot}

\documentclass[10pt,onecolumn]{article}

\makeatletter
\if@twocolumn
  \usepackage[dvips,letterpaper,top=0.5in, bottom=0.5in, left=0.75in, right=0.5in,includefoot,heightrounded]{geometry}
\else
  \usepackage[dvips,letterpaper,margin=1in,includefoot,heightrounded]{geometry}
\fi

\usepackage{srcltx}

\usepackage{amsmath}
\usepackage{amssymb,amsfonts}

\usepackage{abstract}

\usepackage{epsfig}
\usepackage[usenames,dvipsnames]{color}
\usepackage[usenames,dvipsnames]{xcolor}
\usepackage{subfigure}

\usepackage[para]{threeparttable}

\usepackage{booktabs}

\usepackage{setspace}
\usepackage{flushend}
\usepackage{multicol}

\usepackage{cite}
\usepackage{url}
\usepackage[normalem]{ulem}
\usepackage{bm}

\usepackage{enumerate}

\usepackage{multirow}
\usepackage{array}

\usepackage{hyperref}

\usepackage[sc,tiny,center]{titlesec}
       \usepackage{mathptmx}

\newtheorem{definition}{Definition}

\newcommand{\getA}[1]{{#1}^{(a)}}
\newcommand{\getB}[1]{{#1}^{(b)}}
\newcommand{\getC}[1]{{#1}^{(c)}}
\newcommand{\getD}[1]{{#1}^{(d)}}
\newcommand{\getQ}[1]{{#1}^{(q)}}
\newcommand{\getP}[1]{{#1}^{(p)}}



\title{%
A Row-parallel 8$\times$8 \mbox{2-D} DCT Architecture
Using Algebraic Integer Based Exact Computation
}

\author{%
A.~Madanayake%
\thanks{%
A.~Madanayake, N.~T.~Rajapaksha and A.~Edirisuriya
are with the
Department of Electrical and Computer Engineering,
University of Akron, Akron, OH, USA
Email: \url{arjuna@uakron.edu}
}
\and
R.~J.~Cintra%
\thanks{%
R.~J.~Cintra is with 
the Signal Processing Group,
Departamento de Estat\'{\i}stica, 
Universidade Federal de Pernambuco.
E-mail: 
\protect\url{rjdsc@stat.ufpe.org}
}
\and
D.~Onen%
\thanks{%
D.~Onen, V.~S.~Dimitrov and L.~T.~Bruton
are with
Department of Electrical and Computer Engineering, 
University of Calgary, Calgary, AB, Canada.
}
\and
V.~S.~Dimitrov${}^\ddagger$
\and
N.~T.~Rajapaksha${}^\ast$
\and
L.~T.~Bruton${}^\ddagger$
\and
A.~Edirisuriya${}^\ast$
}

\date{}

\newcommand{\myabstract}{%
An algebraic integer (AI) based time-multiplexed row-parallel architecture and two final-reconstruction step (FRS) algorithms are proposed for
the implementation of bivariate AI-encoded
2-D discrete cosine
transform (DCT).
The architecture directly realizes an error-free \mbox{2-D}
DCT without using FRSs between row-column transforms, leading to an 8$\times$8
2-D DCT which is \emph{entirely free of quantization errors} in AI basis.
As a result, the user-selectable accuracy for each of the coefficients in the FRS  facilitates each of the 64~coefficients to have its precision
set independently of others, avoiding
the leakage of quantization noise
between channels as is the case for published DCT designs.
The proposed FRS uses two approaches based on 
(i)~optimized Dempster-Macleod multipliers and 
(ii)~expansion factor scaling.
This architecture enables low-noise high-dynamic range applications in
digital video processing that 
requires full control of the finite-precision computation of the \mbox{2-D} DCT.
The proposed architectures and FRS techniques are experimentally verified and validated using hardware implementations that are physically realized and verified on FPGA chip.
Six designs, for 4- and 8-bit input word sizes, 
using the two proposed FRS schemes, have been
designed, simulated, physically implemented and measured. 
The maximum clock rate and block-rate achieved among 8-bit input designs 
are 
307.787~MHz and 38.47~MHz,
respectively, 
implying a pixel rate of 8$\times$307.787$\approx$2.462~GHz 
if eventually embedded in a real-time video-processing system.
The equivalent frame rate is about 1187.35~Hz for 
the image size of 1920$\times$1080.
All implementations are functional on 
a Xilinx Virtex-6 XC6VLX240T FPGA device.
}

\newcommand{\mykeywords}{%
DCT, Algebraic Integer Quantization, FPGA design
}

\begin{document}

\makeatletter
\if@twocolumn

\twocolumn[%
  \maketitle
  \begin{onecolabstract}
    \myabstract
  \end{onecolabstract}
  \begin{center}
    \small
    \textbf{Keywords}
    \\\medskip
    \mykeywords
  \end{center}
  \bigskip
]
\saythanks

\else

  \maketitle
  \begin{abstract}
    \myabstract
  \end{abstract}
  \begin{center}
    \small
    \textbf{Keywords}
    \\\medskip
    \mykeywords
  \end{center}
  \bigskip
  \onehalfspacing
\fi

\section{Introduction}

High-quality digital video in multimedia devices and video-over-IP networks connected to the Internet are
under exponential growth and therefore
the demand for applications capable of high dynamic range (HDR) video is accordingly increasing.
Some HDR imaging applications include 
automatic surveillance~\cite{huei2009stereoscopic,kao2008fusing,carrillo2009surveillance,surv2}, 
geospatial remote sensing~\cite{geo1}, 
traffic cameras~\cite{traffic1}, 
homeland security~\cite{surv2}, 
satellite based imaging~\cite{satt1,satt2,satt3}, 
unmanned aerial vehicles~\cite{uav1,uav2,uav3}, 
automotive industry~\cite{marsi2007video},
and 
multimedia wireless sensor networks~\cite{wsn1}.
Such HDR video systems operating at high resolutions
require 
an associate hardware capable of significant 
throughput at allowable area-power complexity.

Efficient codec circuits 
capable of both high-speeds of operation \emph{and} high numerical accuracy are needed for next-generation systems.
Such systems may process massive amounts of video feeds, each at high resolution, with minimal noise and distortion while consuming as little energy as possible~\cite{westwaterbook_video_compression}.

The 
two-dimensional (2-D)
discrete cosine transform (DCT) operation
is fundamental to almost all real-time video compression systems. 
The circuit realization of the DCT directly relates to 
noise, 
distortion, 
circuit area, 
and 
power consumption 
of the related video devices~\cite{westwaterbook_video_compression}.
Usually,
the \mbox{2-D} DCT is computed by
successive calls of the
one-dimensional (\mbox{1-D}) DCT applied to the 
columns of an 8$\times$8 sub-image;
then to the rows of the transposed 
resulting intermediate calculation~\cite{suzuki2010integer}.
The VLSI implementation of trigonometric transforms such as DCT and DFT is indeed an active
research area~\cite{madisetti_willson, swamy_chiper,meher_swamy,sung_shieh1, huang_sung_shieh, chen_chang, meher_patra_vinod_2006, meher_unified, nayak, meher_dft, meher_patra_1, tumeo, guo_ju, sun_donner, shams_pan_bayoumi, lin_yu_van,huang_chen_lai}.

An ideal 8-point \mbox{1-D} DCT requires
multiplications by numbers 
in the form
$c[n]=\cos( n \pi / 16 )$,
$n=0,1,\ldots,7$.
These constants impose computational difficulties
in terms of number binary representation
since they are not rational.
Usual DCT implementations adopt a compromise solution 
to this problem
employing
truncation or rounding off~\cite{roundoff1,roundoff2} 
to approximate such quantities.
Thus,
instead of employing the exact value $c[n]$,
a quantized value is considered.
Clearly,
this operation introduces errors.

One way of addressing this problem is to employ 
algebraic integer (AI)
encoding~\cite{DimitrovJullienMiller1998,wahidbook}.
AI-encoding philosophy consists of mapping 
possibly irrational numbers
to array of integers,
which can be arithmetically manipulated without errors.
Also,
depending on the numbers to be encoded,
this mapping can be exact.
For example,
all
8-point DCT multipliers can be given
an exact AI representation~\cite{baghaie2001systolic}.
Eventually,
after computation is performed,
AI-based algorithms require a
final reconstruction step (FRS)
in order
to map the resulting encoded integer arrays
back into usual fixed-point representation
at a given precision~\cite{DimitrovJullienMiller1998}.

Besides the numerical representation issues,
error propagation also plays a role.
In particular,
when considering the fixed-point realization of 
the multiplication operation,
quantization errors are prone to be amplified in the DCT computation~\cite{quant_noise1,quant_noise2}.
Quantization noise at a particular \mbox{2-D} DCT coefficient can have significant correlation with noise in other coefficients depending
on the statistics of the video signal of interest~\cite{quant_noise1,quant_noise2, shams_pan_bayoumi, huang_chen_lai}. 
Combating noise injection, 
noise coupling,
and noise amplification
is a concern
in a practical DCT implementation~\cite{quant_noise1,quant_noise2,roundoff1,roundoff2, shams_pan_bayoumi, huang_chen_lai}.

In~\cite{wahid20042D,wahid2005error},
AI-based procedures for the \mbox{2-D} DCT are proposed.
Their architecture was based on the 
low-complexity Arai algorithm~\cite{arai1988dct},
which formed
the building-block of each \mbox{1-D} DCT using 
AI number representation. The Arai algorithm is a popular algorithm for video and image
processing applications because of its relatively low computational complexity. It is noted that the
8-point Arai algorithm only needs five multiplications 
to generate the eight output coefficients.
Thus, we naturally choose this low complexity algorithm as a foundation for proposing optimized
architectures having lower complexity and lower-noise. 
However,
such design required the algebraically encoded numbers to be
reconstructed to their fixed-point format
by the end of column-wise DCT calculation by means of an intermediate
reconstruction step.
Then data are re-coded to enter into 
the row-wise DCT calculation block~\cite{wahid20042D,wahid2005error}.
This approach is not ideal because it introduces
both numerical representation errors
and error propagation from the intermediate FSR to subsequent blocks.

We propose a digital hardware architecture 
for the $8\times8$ \mbox{2-D} DCT
capable of
(i)~arbitrarily high numeric accuracy
and
(ii)~high-throughput.
To achieve these goals
our design maintains
the signal flow 
free of quantization errors in 
all its intermediate computational steps
by means of a novel doubly AI encoding concept.
No intermediate reconstruction step is introduced
and the entire computation truly occurs
over the AI structure.
This prevents error propagation throughout intermediate computation,
which would otherwise result in error correlation 
among the final DCT coefficients.
Thus
errors are totally confined to a single FRS
that maps the resulting doubly AI encoded DCT coefficients
into fixed-point representations~\cite{DimitrovJullienMiller1998}. 
This procedure allows the selection of individual levels of precision for 
each of the 64~DCT spectral components
at the FRS.
At the same time,
such flexibility
does not affect 
noise levels or 
speed of other sections of the \mbox{2-D} DCT.

This works extends
the 8-point 
\mbox{1-D} AI-based DCT architecture~\cite{wahid20042D,wahid2005error,wahidbook}  
into a fully-parallel time-multiplexed \mbox{2-D} architecture 
for 8$\times$8 data blocks.
The fundamental differences are
(i)~the absence of any intermediate reconstruction step;
(ii)~a new doubly AI encoding scheme;
and
(iii)~the utilization of a single FRS.
The proposed \mbox{2-D} $8\times8$ architecture has the following characteristics:
(i)~independently selectable precision levels for 
the \mbox{2-D} DCT coefficients;
(ii)~total absence of multiplication operations; and
(iii)~absence of leakage of quantization noise between coefficient channels.
The proposed architectures
aim at
performing the FRS operation directly in the bi-variate encoded \mbox{2-D} AI basis.
We introduce designs
based on
(i)~optimized Dempster-Macleod multipliers
and on
(ii)~the expansion factor approach~\cite{britanak2007discrete}.
All hardware implementations 
are designed to be
realized on 
field programmable gate arrays~(FPGAs)
from Xilinx~\cite{xil_web}.

This paper unfolds as follows.
In Section~\ref{section.algebraic}
we review 
existing designs
and 
the main theoretical points
of number representation based on AI.
We keep our focus on the core results
needed for our design.
Section~\ref{section.architecture}
brings a description of
the proposed circuitry and hardware architecture
in block level detail.
In Section~\ref{sec:frs} 
strategies for obtaining the FRS block are proposed and described.
Simulation results and actual test measurements
are reported in Section~\ref{section.test}.
Concluding remarks are drawn in Section~\ref{section.conclusions}.

\section{Review}
\label{section.algebraic}

The AI encoding was originally proposed for digital signal processing systems
by Cozzens and Finkelstein~\cite{cozzens1987range}. 
Since then it has been
adapted for the VLSI implementation of 
the \mbox{1-D} DCT and other trigonometric transforms 
by 
Julien \emph{et al.}~in~\cite{fu_jullien_dimitrov, fu_jullien_dimitrov_ahmadi_miller, fu_soc,fu_2004, fu_presentation}, 
leading to a \mbox{1-D} bivariate encoded Arai DCT algorithm by
Wahid and Dimitrov~\cite{wahid20042D,wahid2005error,wahid_ko,wahidbook}.
Recently, 
subsequent contributions by Wahid \emph{et al.}
(using bivariate encoded \mbox{1-D} Arai DCT blocks for row and column transforms of the \mbox{2-D} DCT) 
has led to practical area-efficient VLSI video processing circuits 
with low-power consumption~\cite{wahid2011, wahid_hindawi, wahid_murtuza}.
We now briefly summarize the state-of-the-art in both~\mbox{1-D} and~\mbox{2-D} DCT VLSI cores based on conventional fixed-point arithmetic 
as well as on AI encoding.

\subsection{Summary and Comparison with Literature}

\subsubsection{Fixed-Point DCT VLSI Circuits}

A unified distributed-arithmetic parallel architecture for the computation of DCT and the DST was proposed in~\cite{meher_unified}. A direct-connected \mbox{3-D} VLSI architecture for the \mbox{2-D} prime-factor DCT that does not need a transpose memory (buffer) is available in~\cite{nayak}. A pioneering implementation at a clock of 100 MHz on {0.8}~$\mu$m CMOS technology for the \mbox{2-D} DCT with block-size $8\times8$ which is suitable for HDTV applications is available in~\cite{madisetti_willson}.

An efficient VLSI linear-array for both $N$-point DCT and IDCT using a subband decomposition algorithm that results in computational- and hardware-complexity of $\mathcal{O}(5N/8)$ with FPGA realization is reported in~\cite{sung_shieh1}. Recently, VLSI linear-array \mbox{2-D} architectures and FPGA realizations having computation complexity $\mathcal{O}(5N/8)$ (for forward DCT) was reported in~\cite{huang_sung_shieh}.

An efficient adder-based \mbox{2-D} DCT core on 0.35~$\mu$m CMOS using cyclic convolution is described in~\cite{guo_ju}. 
A high-performance video transform engine employing a space-time scheduling scheme for computing the \mbox{2-D} DCT in real-time has been proposed and implemented in 0.18~$\mu$m CMOS~\cite{chen_chang}.
A systolic-array algorithm using a memory based design for both the
DCT and the discrete sine transform
which is suitable for real-time VLSI realization 
was proposed in~\cite{swamy_chiper}.
An FPGA-based system-on-chip realization of the \mbox{2-D} DCT for $8\times8$ block size that operates at 107 MHz with a latency of 80~cycles is available in~\cite{tumeo}. 
A low-complexity IP core for quantized $8\times8/4\times4$ DCT 
combined with 
MPEG4 codecs and FPGA synthesis is available in~\cite{sun_donner}. 
``New distributed-arithmetic (NEDA)'' based low-power $8\times8$ \mbox{2-D} DCT is reported in~\cite{shams_pan_bayoumi}. 
A reconfigurable 
processor on TSMC~0.13~$\mu$m CMOS technology operating at 100~MHz is
described in~\cite{lin_yu_van}
for the calculation of the fast Fourier transform and the \mbox{2-D} DCT. 
A high-speed \mbox{2-D} transform architecture based on NEDA technique and having unique kernel for multi-standard video processing is described in~\cite{huang_chen_lai}.

\subsubsection{AI-based DCT VLSI Circuits}

The following AI-based realizations of \mbox{2-D} DCT computation relies on 
the row- and column-wise application of \mbox{1-D} DCT cores that 
employ AI quantization~\cite{fu_jullien_dimitrov, fu_jullien_dimitrov_ahmadi_miller, fu_soc,fu_2004, fu_presentation}. 
The architectures proposed by Wahid \emph{et al.} rely on 
the low-complexity Arai Algorithm 
and 
lead to low-power realizations~\cite{wahid20042D,wahid2005error,wahid2011,wahid_hindawi, wahid_ko}. 
However, 
these realizations also are based on repeated application 
along row and columns of an fundamental \mbox{1-D} DCT building block 
having an FRS section at the output stage. 
Here, 
$8\times8$ \mbox{2-D} DCT refers 
to the use of bivariate encoding in the AI basis 
and 
not to the a true AI-based \mbox{2-D} DCT operation.

A $4\times4$ approximate \mbox{2-D}-DCT using AI quantization is reported in~\cite{nandi}. Both FPGA implementation and ASIC synthesis on 90 nm CMOS results are provided. Although~\cite{nandi} employs AI encoding, it is not an error-free architecture. The low complexity of this architecture makes it suitable for H.264 realizations.

\subsection{Preliminaries for Algebraic Integer Encoding and Decoding}

In order to prevent quantization noise,
we adopt the AI representation.
Such representation is based on
a mapping function that links
input numbers to integer arrays.

This topic is a major and classic field in number theory.
A famous exposition is
due to Hardy and Wright~\cite[Chap.~XI and XIV]{hardy1975numbers},
which is widely regarded as masterpiece on this subject
for its clarity and depth.
Pohst also brings a didactic explanation 
in~\cite{pohst1993computational} with emphasis on
computational realization.
In~\cite[p.~79]{pollard1975theory},
Pollard and Diamond devote an entire chapter to the connections
between
algebraic integers and integral basis.
In the following,
we furnish an overview
focused on the practical aspects of AI,
which may be useful for circuit designers.

\begin{definition}
A real or complex number is called an algebraic integer if it
is a root of a monic polynomial with 
integer coefficients~\cite{hardy1975numbers,baghaie2001systolic}.
\end{definition}

The set of algebraic integers have useful mathematical properties.
For instance,
they form a commutative ring,
which means that addition and multiplication operations
are commutative and also satisfies distribution over addition.

A general AI encoding mapping
has the following format
\begin{align*}
f_\text{enc}(x; \mathbf{z}) = \mathbf{a},
\end{align*}
where 
$\mathbf{a}$ is a multidimensional array of integers
and
$\mathbf{z}$ is a fixed multidimensional array
of algebraic integers.
It can be shown that
there always exist integers such that
any real number can be represented with 
arbitrary precision~\cite{cozzens1987range}. 
Also there are real numbers that can be represented \emph{without} error.

Decoding operation is furnished by
\begin{align*}
f_\text{dec}(\mathbf{a}; \mathbf{z})
=
\mathbf{a} \bullet \mathbf{z}
,
\end{align*}
where
the binary operation $\bullet$ is the generalized inner product ---
a component-wise inner product of multidimensional arrays.
The elements of~$\mathbf{z}$ constitute the AI basis.
In hardware,
decoding is often performed by 
an FRS block,
where the AI basis $\mathbf{z}$ is
represented as precisely as required.

As an example,
let the AI basis be such that
$\mathbf{z} = \begin{bmatrix}1 & z_1 \end{bmatrix}^T$, 
where $z_1$ is the algebraic integer $\sqrt{2}$
and the superscript~${}^T$ denotes the transposition operation.
Thus, a possible AI encoding mapping is
$f_\text{enc}(x; \mathbf{z}) = 
\mathbf{a} = \begin{bmatrix}a_0 & a_1\end{bmatrix}^T$,
where~$a_0$ and $a_1$ are integers.
Encoded numbers are then represented by a 2-point vector of integers.
Decoding operation is simply given by the usual inner product:
$x = \mathbf{a} \bullet \mathbf{z} = a_0 + a_1 z_1$.
For example, the number $1-2\sqrt{2}$ has the following encoding:
\begin{align*}
f_\text{enc}
\left(
1-2\sqrt{2} ; 
\begin{bmatrix}1 \\ \sqrt{2}\end{bmatrix}
\right) 
=
\begin{bmatrix}1 \\ -2\end{bmatrix},
\end{align*}
which is an \emph{exact} representation.

In principle,
any number can be represented in 
an arbitrarily high precision~\cite{cozzens1987range,cozzens1985computing}.
However,
within a limited dynamic range for the employed integers,
not all numbers can be exactly encoded.
For instance, considering the real number $\sqrt{3}$,
we have 
$f_\text{enc}(\sqrt{3}; \begin{bmatrix}1& \sqrt{2}\end{bmatrix}^T) = \begin{bmatrix}88 & -61\end{bmatrix}^T$,
where integers were limited to be 8-bit long.
Although very close,
the representation is not exact:
\begin{align*}
f_\text{dec}
\left(
\begin{bmatrix}88 \\ -61\end{bmatrix}; 
\begin{bmatrix}1 \\ \sqrt{2}\end{bmatrix}
\right) 
- \sqrt{3}
\approx 9.21 \times 10^{-4}.
\end{align*}

In a similar way,
the multipliers required by the DCT
could be encoded into 2-point integer vectors:
$f_\text{enc}(c[n]; \mathbf{z}) = \begin{bmatrix}a_0[n] & a_1[n]\end{bmatrix}^T$.
Given that the DCT constants are algebraic integers~\cite{baghaie2001systolic},
an exact AI representation can be derived~\cite{ard1}.
Thus,
the integer sequences $a_0[n]$ and $a_1[n]$ can be easily 
realized in VLSI hardware.

The multiplication between two numbers represented
over an AI basis may be interpreted
as a modular polynomial multiplication
with respect to the monic polynomial that
defines the AI basis.
In the above particular illustrative example,
consider the multiplication of the following
pair of numbers
$a_0 + a_1 z_1$ with $b_0 + b_1 z_1$, where $b_0$ and $b_1$ are integers.
This operation is equivalent to the computation of
the following expression:
\begin{align*}
(a_0 + a_1 x)
\cdot
(b_0 + b_1 x)
\pmod{x^2-2}
.
\end{align*}
Thus,
existing
algorithms for fast polynomial multiplication 
may be of consideration~\cite[p.~311]{blahut2010fast}.

In practical terms,
a good AI representation
possesses a basis such that:
(i)~the required constants can be represented \emph{without} error;
(ii)~the integer elements provided by the representation are 
sufficiently small to allow a simple architecture design and
fast signal processing;
and
(iii)~the basis itself contains few elements to facilitate
simple encoding-decoding operations.

Other AI procedures allow the constants to be approximated,
yielding much better options for encoding, at the cost of introducing error within the transform (before the FRS)~\cite{baghaie2001systolic}.

\subsection{Bivariate AI Encoding}

Depending on the
DCT algorithm employed,
only the cosine of a few arcs are in fact required.
We adopted the Arai DCT algorithm~\cite{arai1988dct};
and the required elements for this particular
\mbox{1-D} DCT method are only~\cite{wahid20042D,wahid2005error,wahidbook}:
\begin{align*}
c[4] &= \cos \frac{4\pi}{16}, \quad
c[6] = \cos \frac{6\pi}{16},
\\
c[2]-c[6] &= \cos \frac{2\pi}{16} - \cos \frac{6\pi}{16}, 
\\
c[2]+c[6] &= \cos \frac{2\pi}{16} + \cos \frac{6\pi}{16}.
\end{align*}

These particular values can be conveniently
encoded as follows.
Considering 
$z_1 = \sqrt{2+\sqrt{2}} + \sqrt{2-\sqrt{2}}$
and
$z_2 = \sqrt{2+\sqrt{2}} - \sqrt{2-\sqrt{2}}$,
we adopt the following \mbox{2-D} array
for AI encoding:
\begin{align*}
\mathbf{z}
=
\begin{bmatrix}
1 & z_1 \\
z_2 & z_1z_2
\end{bmatrix}
.
\end{align*}
This leads to a \mbox{2-D} encoded coefficients of the form (scaled by 4):
\begin{align*}
f_\text{enc}(x; \mathbf{z}) 
= 
\mathbf{a}
=
\begin{bmatrix}
a_{0,0} & a_{1,0} \\ 
a_{0,1} & a_{1,1} 
\end{bmatrix}
.
\end{align*}
Such encoding is referred to as bivariate.
For this specific AI basis,
the required cosine values
possess
an error-free and sparse representation
as given in Table~\ref{table.2}~\cite{wahid2005error,wahid20042D,wahidbook}.
Also we note that this representation utilizes
very small integers
and therefore is
suitable for fast arithmetic computation.
Moreover, 
these employed integers are powers of two, 
which
require 
no hardware components other than wired-shifts,
being cost-free.

\begin{table}
\centering
\caption{\mbox{2-D} AI encoding of Arai DCT constants}
\label{table.2}
\begin{tabular}{ccccc}
\hline
\noalign{\smallskip}
$c[4]$ & $c[6]$ & $c[2]-c[6]$  & $c[2]+c[6]$ \\ 
\hline
\noalign{\smallskip}
$\begin{bmatrix}0 & 0 \\ 0 & 1 \end{bmatrix}$
&
$\begin{bmatrix}0 & 1 \\ -1 & 0 \end{bmatrix}$
&
$\begin{bmatrix}0 & 0 \\ 2 & 0 \end{bmatrix}$
&
$\begin{bmatrix}0 & 2 \\ 0 & 0 \end{bmatrix}$
\\
\noalign{\smallskip}
\hline
\end{tabular}
\end{table}

Encoding an arbitrary real number can be a sophisticated operation
requiring the usage of look-up tables
and greedy algorithms~\cite{dimitrov1997eisenstein}.
Essentially,
an exhaustive search is required 
to obtain the most accurate representation.
However,
integer numbers can be encoded effortlessly:
\begin{align}
\label{eq.trivial.coding}
f_\text{enc}(m; \mathbf{z})
=
\begin{bmatrix}
m & 0 \\
0 & 0
\end{bmatrix},
\end{align}
where $m$ is an integer.
In this case,
the encoding step is unnecessary.
Our proposed design takes advantage of this property.

For a given encoded number $\mathbf{a}$,
the decoding operation is simply expressed by:
\begin{align*}
f_\text{dec}(\mathbf{a}; \mathbf{z})
=
\mathbf{a} \bullet \mathbf{z}
=
a_{0,0} + a_{1,0} z_1 + 
a_{0,1} z_2 + a_{1,1} z_1 z_2
.
\end{align*}
In terms of circuitry design,
this operation is usually performed
by the FRS.

In order to reduce and simplify the employed notation,	
hereafter a superscript notation is used
for identifying the bivariate AI encoded coefficients.
For a given real $x$,
we have	the following representation
\begin{align}
\label{z4.representation}
\begin{bmatrix}
\getA{x} & \getB{x} \\ 
\getC{x} & \getD{x}
\end{bmatrix}	
\equiv
x
=
\getA{x} 
+
\getB{x} z_1
+
\getC{x} z_2
+
\getD{x} z_1 z_2,	
\end{align}
where superscripts $\getA{}$, $\getB{}$, $\getC{}$, and $\getD{}$
indicate the encoded integers associated to
basis elements 1, $z_1$, $z_2$, and $z_1z_2$,
respectively.
We denote this basis as
$\mathbf{z}_4 = \{ 1,z_1,z_2,z_1z_2\}$.

It is worth to emphasize that in the \mbox{2-D} AI encoding
the
equivalence between 
the algebraic integer multiplication and
the polynomial modular multiplication 
does not hold true.
Thus,
a tailored computational technique to handle this operation
must be developed.

\section{\mbox{2-D} AI DCT Architecture}
\label{section.architecture}

An 8$\times$8 image block $\mathbf{A}$ has its
\mbox{2-D} DCT transform mathematically expressed by~\cite{suzuki2010integer}:
\begin{align}
\label{equation.2d.dct}
\left( \mathbf{C} \cdot (\mathbf{C} \cdot \mathbf{A})^T  \right)^T,
\end{align}
where $\mathbf{C}$ is the usual DCT matrix~\cite{britanak2007discrete}.
It is straightforward to notice that this operation
corresponds to
the column-wise application of the \mbox{1-D} DCT to the
input image $\mathbf{A}$,
followed by a transposition,
and then
the row-wise application of the \mbox{1-D} DCT to the resulted matrix.

The \mbox{2-D} DCT realizations 
in~\cite{wahidDimitrov2005,mayer-base2001optimal,wahid2005error,wahid20042D} 
use the AI encoding scheme with
decoding sections placed in between 
the row- and column-wise \mbox{1-D} DCT operations. 
This intermediate reconstruction step leads to 
the introduction of 
quantization noise 
and 
cross-coupling of correlated noise components.
In contrast,
we employ 
a bivariate
AI encoding,
maintaining the computation over AI arithmetic
to completely avoid arithmetic errors 
within the algorithm ~\cite{ard1}.

\begin{figure}
\centering
\includegraphics[scale=0.14]{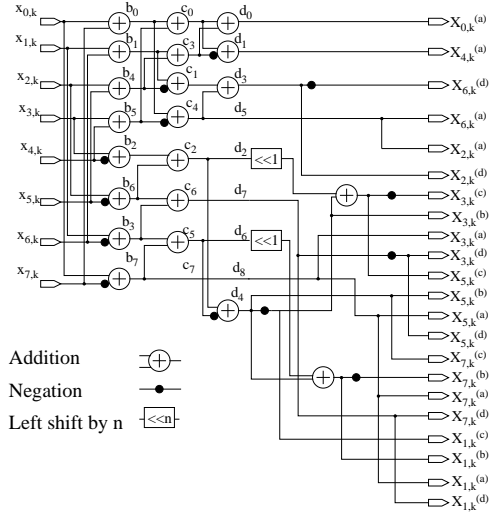}
\caption{1-D AI Arai DCT block used in Fig.~\ref{fig1}\cite{wahid20042D}. }
\label{fig:AI_DCT_block}
\end{figure}

\begin{figure}
\centering
\hspace{-2mm}
\input{rjdsc-AI-TB.pstex_t}
\caption{1-D AI transpose buffer used in Fig.~\ref{fig1}.}
\label{fig:AI_TB}
\end{figure}
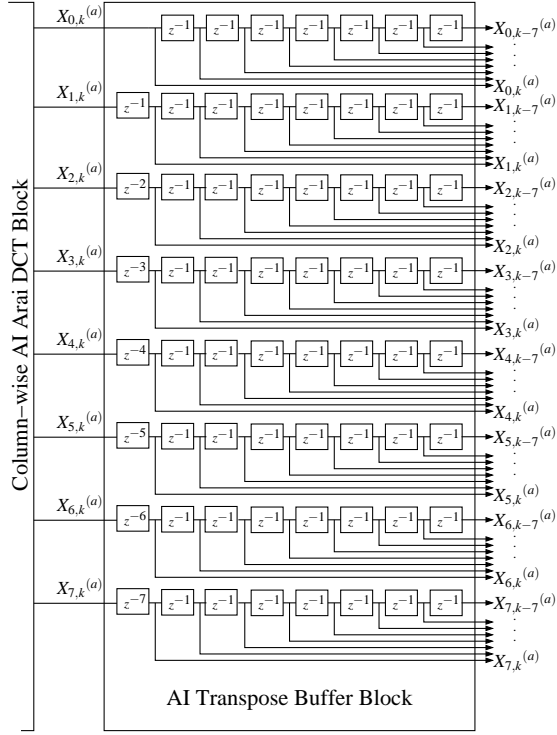

\begin{figure*}
\centering
\input{rjdsc-fig1.pstex_t}
\caption{The \mbox{2-D} AI-DCT consists of
an input section having a decimation structure,
\mbox{1-D} 8-point AI-DCT block for column-wise DCTs,
a real-time AI-TB,
four parallel \mbox{1-D} 8-point AI-DCT blocks for row-wise DCTs,
and an FRS ~\cite{ard1}.}
\label{fig1}
\end{figure*}
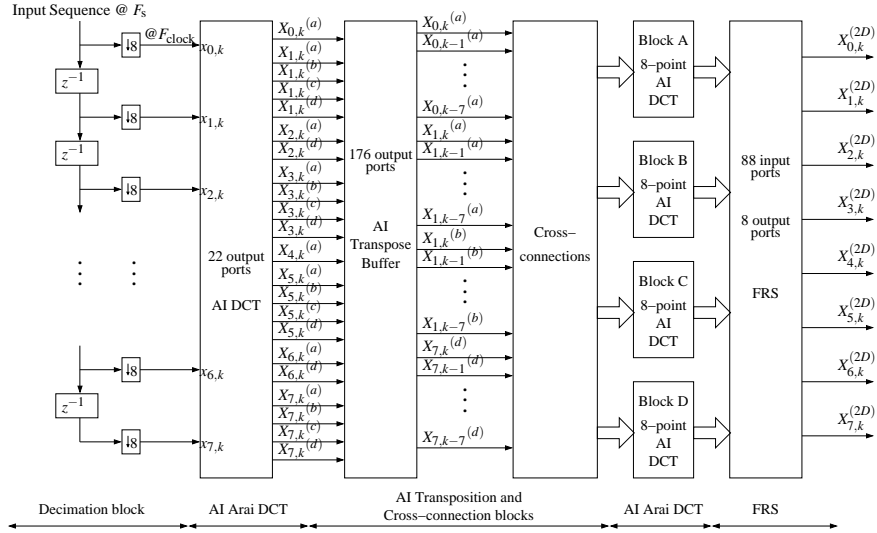

The proposed architecture
consists of five sub-circuits~\cite{ard1}:
(i)~an input decimator circuit;
(ii)~an 8-point AI-encoded \mbox{1-D} DCT block shown in Fig.~\ref{fig:AI_DCT_block} 
which performs column-wise computation based on
the Arai algorithm~\cite{arai1988dct} and furnishes the intermediate result
$\mathbf{C} \cdot \mathbf{A}$ 
in the AI domain;
(iii)~an AI-based transposition buffer shown in Fig.~\ref{fig:AI_TB} with 
a wired cross-connection block for obtaining 
$(\mathbf{C} \cdot \mathbf{A})^T$;
(iv)~four parallel instantiations of the same 8-point AI-based Arai DCT block in Fig.~\ref{fig:AI_DCT_block} 
for row-wise computation of eight \mbox{1-D} DCTs,
which results in $\mathbf{C} \cdot (\mathbf{C} \cdot \mathbf{A})^T$;
and
(v)~an FRS circuit for mapping the AI-encoded \mbox{2-D} DCT coefficients 
to 2's complement format.
The last transposition (\ref{equation.2d.dct})
is obtained via wired cross-connections.
The proposed architecture is shown in Fig.~\ref{fig1}.

Our implementation covers items (ii)--(v) listed above.
We now describe in detail 
each of the system blocks.

\subsection{Bit Serial Data Input, SerDes, and Decimation}

We assume that the input video data, in raster-scanned format, 
has already been split into 
8$\times$8 pixel blocks.
We further assume that these blocks can be stacked to form an 
8-column and ($8\times(\text{number of blocks})$)-row
data structure. 
This leads to so-called ``blocked'' video frames, 
each of size 8$\times$8 pixels.
The blocking procedure leads to
a raster-scanned sequence of pixel intensity (or color) values $x_{i,n}$,
$i=0,1,\ldots,7$, 
$n=0, 1, \ldots, 8\times(\text{number of blocks})-1$,
from an 8$\times$8 blocked image.
Notice that we use column-row order for the indexes,
instead of row-column.
Due to the 8$\times$8 size of the \mbox{2-D} DCT computation,
we find it quite convenient to consider the time index $n$ after 
a modular operation
$k\equiv n \pmod{8}$.
Hereafter,
we will refer to the time index as
a modular quantity
$k=0,1,\ldots,7,0,1,\ldots,7,0,1\ldots,7,\ldots$.

The video signal is serially streamed through 
the input port of the architecture 
at a rate of $F_{\text{s}}$. 
A bit serial port connected to a serializer/deserializer (SerDes) 
is required
to be fed using
a bit rate of $8 \times F_{\text{s}}$ without considering overheads. 
As an aside, 
we note that this input bit stream may be typically derived from
optical fiber transmission
or
high throughput Ethernet ports driven at~9.6~Gbps.
Following the SerDes, 
a decimation block converts the input byte sequence into
a row structure by means of delaying and 
downsampling by eight
as shown in Fig.~\ref{fig1}.

Therefore,
the raster-scanned  input is decimated 
in time into eight parallel streams operating rate of 
$F_{\text{clock}}=F_{\text{s}}/8$; 
resulting in eight columns of the input block.
It is important to emphasize that such input data consist
of integer values.
Thus,
they are AI coded without any computation
as shown in (\ref{eq.trivial.coding}).
The obtained column data is submitted to
the column-wise application of the AI-based \mbox{1-D} DCT.

\subsection{An 8-point AI-Encoded Arai DCT Core}

The column-wise transform operation is performed according to 
the 8-point AI-based Arai DCT hardware cores
as designed in~\cite{wahid20042D,wahid2005error} shown in Fig.~\ref{fig:AI_DCT_block}.
Here,
this scheme is employed with the \emph{removal} of its original FRS.
The proposed \mbox{2-D} architecture 
employs an integer arithmetic
entirely defined
over
the AI basis $\mathbf{z}_4$.
This transformation step 
operates at the reduced clock rate of $F_{\text{clock}}$.

Indeed, 
the resulting AI encoded data components
are split in four channels according to
their $\mathbf{z}_4$ basis representation ~\cite{ard1}.
Such outputs are
time-multiplexed mixed-domain partially computed spectral components.
We denote them as
$\getA{X_{i,k}}$, $\getB{X_{i,k}}$, $\getC{X_{i,k}}$, $\getD{X_{i,k}}$,
where 
$i=0,1,\ldots,7$ is the column index
and 
$k$ is the modular time index 
containing the information of the row number.

In hardware, 
this means that the
AI representation
is contained in at most four parallel integer channels ~\cite{ard1}.
Some quantities are known beforehand
to require less than four AI encoded integers
(cf.~(\ref{z4.representation})).
Thus,
in some cases,
less than four connections
are required.
These channels are routed to 
the proposed AI-based transpose buffer (AI-TB) shown in Fig.~\ref{fig:AI_TB},
as
a necessary pre-processing for the subsequent row-wise DCT calculation.

\subsection{Real-time AI-based Transpose Buffer}

Each partially computed transform component 
$\getQ{X_{i,k}}$, $q\in\{ a, b, c, d\}$,
from the column-wise DCT block is 
represented in~$\mathbf{z}_4$.
Such encoded components are stored 
in the proposed AI-TB (shown in Fig.~\ref{fig:AI_TB} only for channel $^{(a)}$),
which computes an 8$\times$8 matrix transposition operation 
in real-time every eight clock cycles.

The proposed AI-TB consists of 
a chain of clocked first-in-first-out (FIFO) buffers 
for each AI-based channel of each component 
of 
the column-wise transformation~\cite{ard1}.
For each parallel integer channel $q$,
there are eight FIFO taps 
clocked at rate~$F_{\text{clock}}$.
Therefore,
the set of FIFO buffers 
leads to $22 \times 8 = 176$ output ports 
from the FIFO buffer section. 
Hard wired cross-connections are used   
that physically realize the required transpose matrix
for the next row-wise DCT section.
These physical connections are encapsulated 
in the cross-connection block in Fig.~\ref{fig1} for brevity.
The AI-TB is clocked at a rate of $F_{\text{clock}}$ 
and yields a new 8$\times$8 block of transposed data
every 64~clock periods of the master clock $F_{\text{s}}$. 
Subsequently,
the transposed AI-encoded 
elements are submitted to four \mbox{1-D} AI DCT cores
operating in parallel.

\subsection{Row-wise DCT Computation}

After route cross-connection,
the output taps from the transposition operation
are 
connected to 32 parallel \mbox{8:1} multiplexers.
Each multiplexer commutes continuously
and 
routes each partially computed DCT component
by cycling through its 3-bit control codes such that 
the $q$ channel inputs 
of each of the four row-wise AI-based DCT cores are provided with
a new set of valid input vectors at rate $F_{\text{clock}}$.

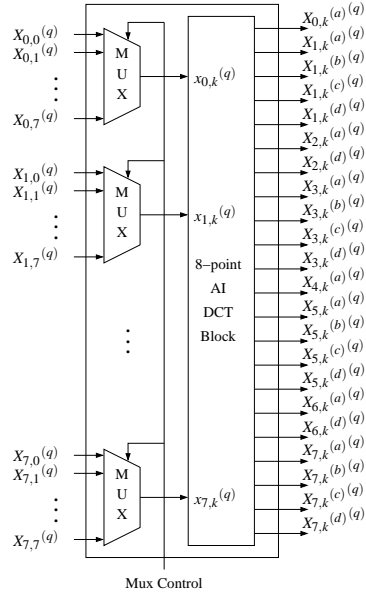
\begin{figure}
\centering
\input{rjdsc-fig2.pstex_t}
\caption{Row-wise DCT block that leads to the \mbox{2-D} DCT of 
the 8$\times$8 input blocks.}
\label{fig2}
\end{figure}

The cores are set in parallel
being able to compute
an 8-point DCT every eight clock cycles of the master clock signal.
This operation performs the required row-wise DCT computation in order
to complete the \mbox{2-D} DCT evaluation,
resulting in a doubly encoded AI representation
$\getP{\getQ{X_{i,k}}}$,
$p,q \in \{a,b,c,d\}$.
Fig.~\ref{fig2} shows the above described block.

\subsection{Final Reconstruction Step}

The output channels for the 64 \mbox{2-D} DCT coefficients are passed through 
the proposed FRS 
for decoding the AI-encoded numbers
back into their fixed-point, binary representation, in~2's complement format. 
Two different architectures are proposed for the FRS.

\section{Final Reconstruction Step}
\label{sec:frs}

The proposed FRS architectures differ from the one in~\cite{wahidDimitrov2005} 
by having individualized circuits to compute each output value
at possibly different precisions.

Indeed,
no FRS circuits are employed in any intermediate~\mbox{1-D} DCT block.
This prevents quantization noise cross-coupling between DCT channels.
Any quantization noise is injected only at the final output.
Therefore noise signals are uncorrelated, 
which further allows the noise for each output
to be independently adjustable and made as low as required. 

\subsection{FRS based on Dempster-Macleod method}
In this method the doubly encoded elements can be decoded according to:
\begin{equation}
\label{final.reconstruction}
\begin{split}
\getQ{X_{i,k}}
=&
\getA{\getQ{X_{i,k}}}
+
\getB{\getQ{X_{i,k}}}
z_1
+
\getC{\getQ{X_{i,k}}}
z_2
+
\\
&
\getD{\getQ{X_{i,k}}}
z_1z_2
,
\quad
q\in\{a,b,c,d\},
\end{split}
\end{equation}
which are then submitted to~(\ref{z4.representation}).
The result is the $k$th row of the
final \mbox{2-D} DCT data
$X_{i,k}$, $i=0,1,\ldots,7$.

Therefore,
for each $q$,
(\ref{final.reconstruction})
unfolds into a particular mathematical expression
as shown below:
\begin{equation}
\label{eq-unfold-a}
\begin{split}
\getA{X_{i,k}}
=&
\getA{\getA{X_{i,k}}} + 
\getB{\getA{X_{i,k}}} z_1 + 
\\
&
\getC{\getA{X_{i,k}}} z_2 + 
\getD{\getA{X_{i,k}}} z_1 z_2,
\end{split}
\end{equation}
\begin{equation}
\begin{split}
\getB{X_{i,k}} z_1
=&
\getA{\getB{X_{i,k}}} z_1 + 
\getB{\getB{X_{i,k}}} z_1^2 + 
\\
&
\getC{\getB{X_{i,k}}} z_1z_2 + 
\getD{\getB{X_{i,k}}} z_1^2 z_2,
\end{split}
\end{equation}
\begin{equation}
\begin{split}
\getC{X_{i,k}} z_2
=&
\getA{\getC{X_{i,k}}} z_2 + 
\getB{\getC{X_{i,k}}} z_1 z_2 + 
\\
&
\getC{\getC{X_{i,k}}} z_2^2 + 
\getD{\getC{X_{i,k}}} z_1 z_2^2,
\end{split}
\end{equation}
\begin{equation}
\label{eq-unfold-d}
\begin{split}
\getD{X_{i,k}} z_1 z_2
=&
\getA{\getD{X_{i,k}}} z_1 z_2 + 
\getB{\getD{X_{i,k}}} z_1^2 z_2 + 
\\
&
\getC{\getD{X_{i,k}}} z_1 z_2^2 + 
\getD{\getD{X_{i,k}}} z_1^2 z_2^2.
\end{split}
\end{equation}
The summation of above quantities returns $X_{i,k}$ (cf.~(\ref{z4.representation})).
Terms depending on $z_1$ and $z_2$
may not be rational numbers.
Indeed, 
they are given by
\begin{equation}
\label{exact-values-of-z}
\begin{split}
z_1 &= \sqrt{2 + \sqrt{2}} + \sqrt{2 - \sqrt{2}} = 2.613125929752\ldots \\
z_2 &= \sqrt{2 + \sqrt{2}} - \sqrt{2 - \sqrt{2}} = 1.082392200292\ldots \\
z_1^2 &= 4+2\sqrt{2} = 6.828427124746\ldots \\
z_2^2 &= 4-2\sqrt{2} = 1.171572875253\ldots \\
z_1z_2 &= 2 \sqrt{2} = 2.82842712474619\ldots \\
z_1z_2^2 &= 4 \sqrt{2 - \sqrt{2}}= 3.061467458920\ldots \\
z_1^2z_2 &= 4 \sqrt{2 + \sqrt{2}} = 7.391036260090\ldots \\
z_1^2z_2^2 &= 8
.
\end{split}
\end{equation}
Multiplier $z_1^2z_2^2=8$ is a power of two and 
can be represented exactly.
Remaining constants require a binary approximation.

Closest signed 12-bit approximations can be employed
to approximate
the above listed numbers.
Such approach furnished the 
quantities below:
\begin{align*}
\widetilde{z_1} &= \frac{669}{2^8} = 2.61328125, & 
\widetilde{z_2} &= \frac{2217}{2^{11}} = 1.08251953125,
\\
\widetilde{z_1^2} &= \frac{437}{2^6} = 6.828125, & 
\widetilde{z_2^2} &= \frac{2399}{2^{11}} = 1.17138671875,
\\
\widetilde{z_1z_2} &= \frac{181}{2^{6}} = 2.828125, & 
\widetilde{z_1z_2^2} &= \frac{3135}{2^{10}} = 3.0615234375,
\\
\widetilde{z_1^2z_2} &= \frac{473}{2^6} = 7.390625.
&&
\end{align*}

Consequently,
the
12-bit approximation
expressions 
related to
$\getQ{X_{i,k}}$ are given by:
\begin{equation}
\label{eq.frs.a}
\begin{split}
\getA{X_{i,k}}
\approx&
\getA{\getA{X_{i,k}}} + 
\frac{669}{2^8} \cdot \getB{\getA{X_{i,k}}}  + 
\\
&
\frac{2217}{2^{11}} \cdot \getC{\getA{X_{i,k}}} + 
\frac{181}{2^{6}} \cdot \getD{\getA{X_{i,k}}}, 
\end{split}
\end{equation}

\begin{equation}
\label{eq.frs.b}
\begin{split}
\getB{X_{i,k}} z_1
\approx&
\frac{669}{2^8} \cdot \getA{\getB{X_{i,k}}} + 
\frac{437}{2^6} \cdot \getB{\getB{X_{i,k}}} + 
\\
&
\frac{181}{2^{6}} \cdot \getC{\getB{X_{i,k}}} + 
\frac{473}{2^6} \cdot \getD{\getB{X_{i,k}}},
\end{split}
\end{equation}
\begin{equation}
\label{eq.frs.c}
\begin{split}
\getC{X_{i,k}} z_2
\approx&
\frac{2217}{2^{11}} \cdot \getA{\getC{X_{i,k}}} + 
\frac{181}{2^{6}} \cdot \getB{\getC{X_{i,k}}} + 
\\
&
\frac{2399}{2^{11}} \cdot \getC{\getC{X_{i,k}}} + 
\frac{3135}{2^{10}} \cdot \getD{\getC{X_{i,k}}},
\end{split}
\end{equation}
\begin{equation}
\label{eq.frs.d}
\begin{split}
\getD{X_{i,k}} z_1 z_2
\approx&
\frac{181}{2^{6}} \cdot \getA{\getD{X_{i,k}}} + 
\frac{473}{2^6} \cdot \getB{\getD{X_{i,k}}} + 
\\
&
\frac{3135}{2^{10}} \cdot \getC{\getD{X_{i,k}}} + 
8 \cdot \getD{\getD{X_{i,k}}}.
\end{split}
\end{equation}

\begin{figure*}
\centering
\subfigure[]{\epsfig{file=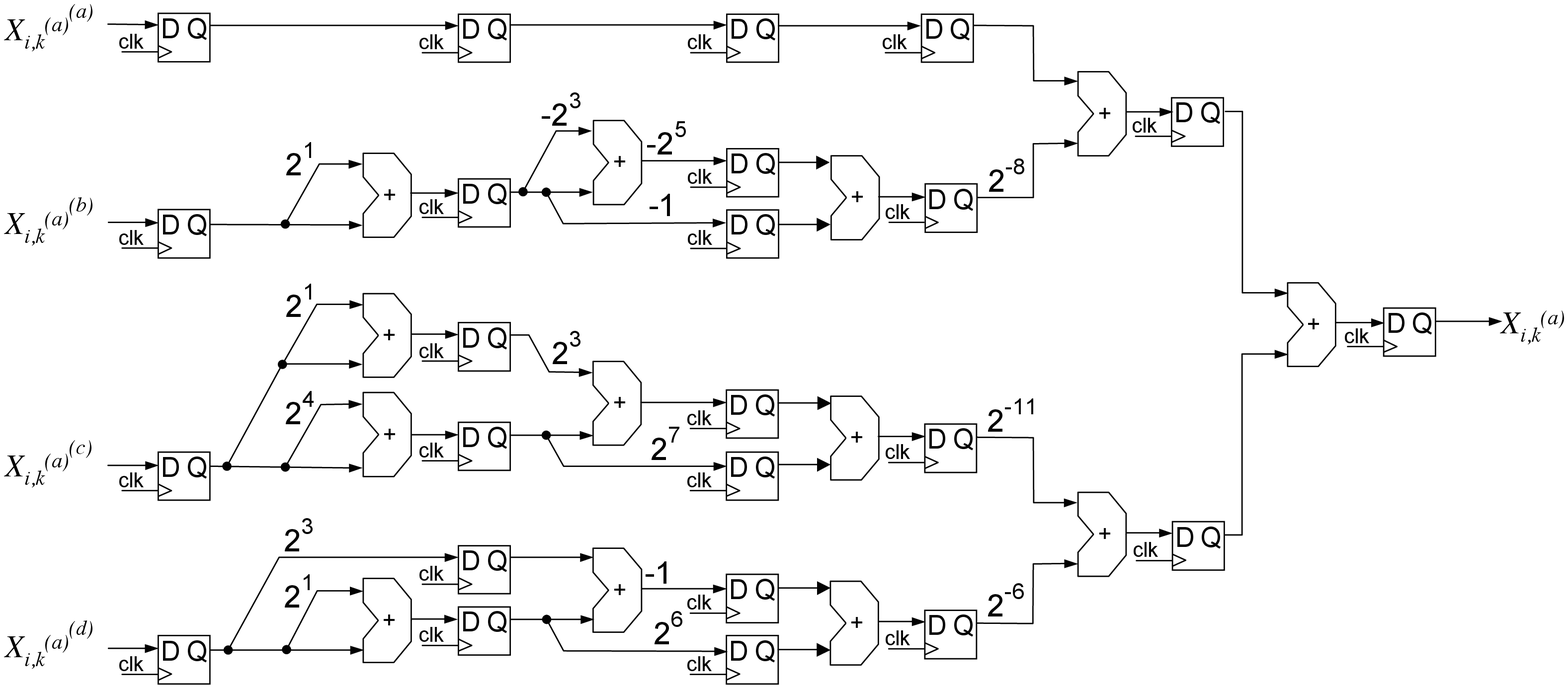,width=0.6\textwidth}}
\subfigure[]{\epsfig{file=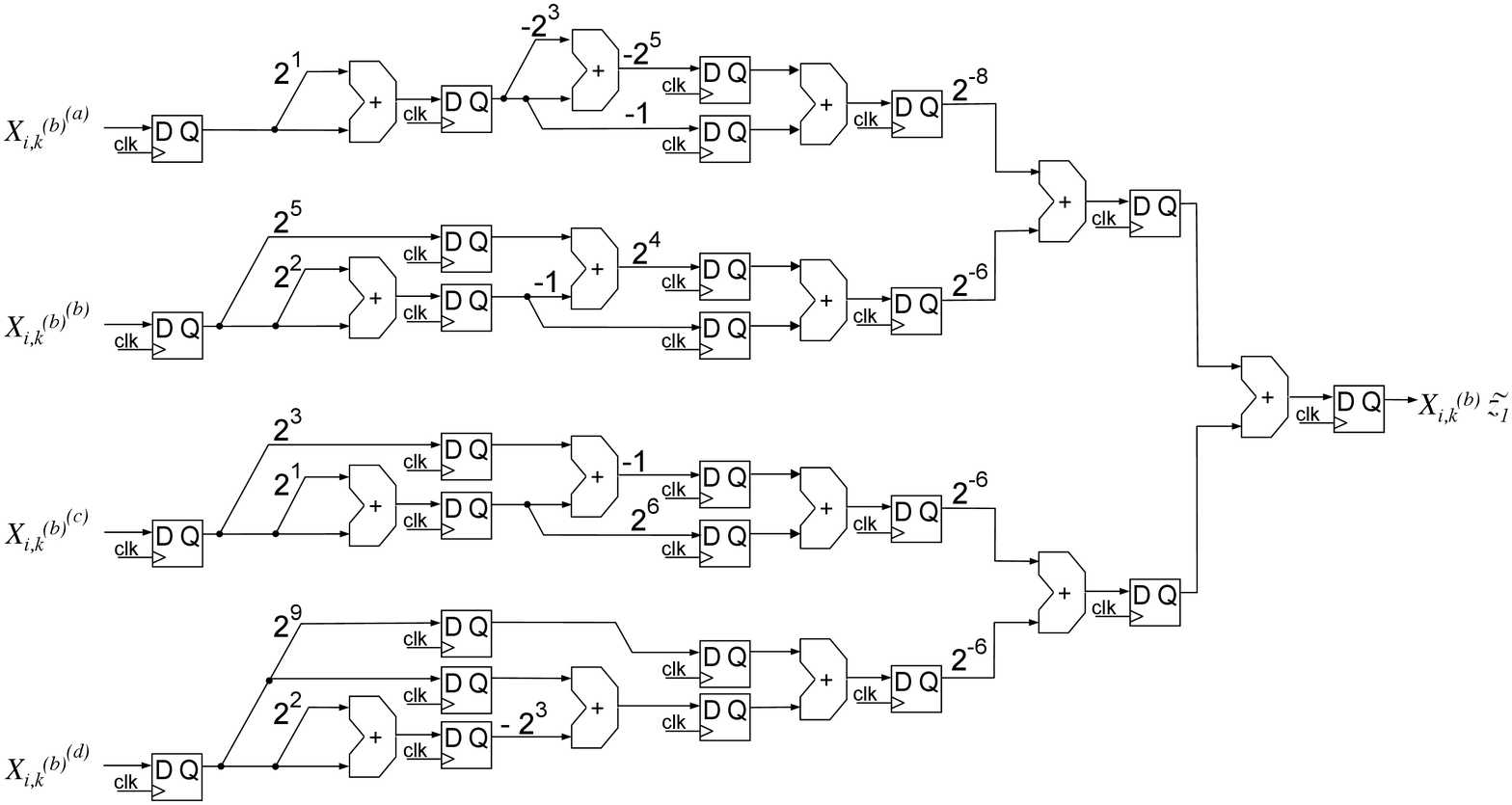,width=0.6\textwidth}}
\caption{Final reconstruction step blocks 
with multi-level pipelining, 
for (\ref{eq.frs.a}) and (\ref{eq.frs.b}), 
respectively.}
\label{fig.frs.1.2}
\end{figure*}

\begin{figure*}
\centering
\subfigure[]{\epsfig{file=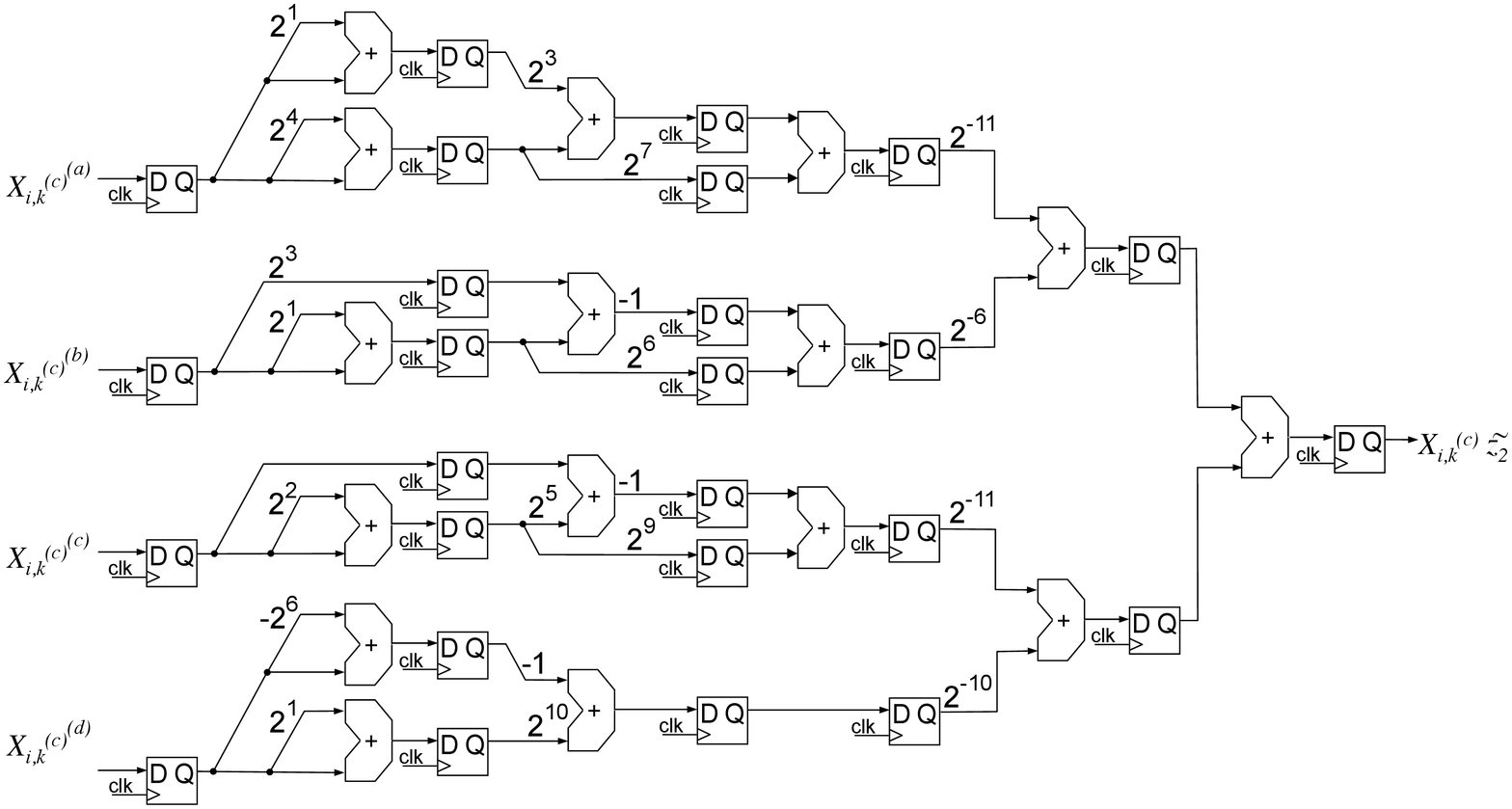,width=0.6\textwidth}}
\subfigure[]{\epsfig{file=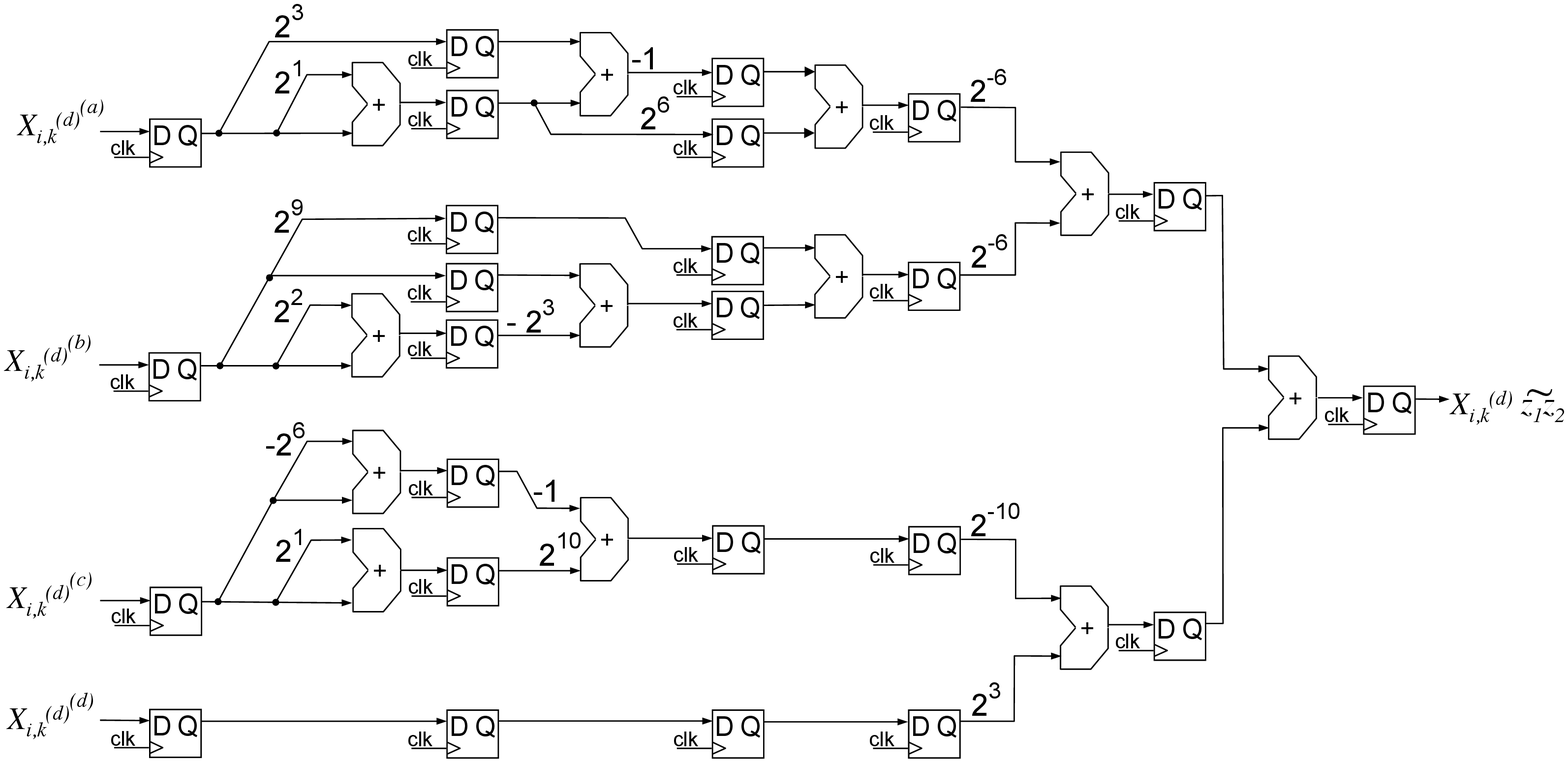,width=0.6\textwidth}}
\caption{Final reconstruction step blocks 
with multi-level pipelining, 
for (\ref{eq.frs.c}) and (\ref{eq.frs.d}), 
respectively.}
\label{fig.frs.3.4}
\end{figure*}

\begin{table}
\centering
\caption{Fast algorithms for required integer multipliers}
\label{table.fast.multiplication}
\begin{tabular}{cl}
\hline 
$m$ & Input: $x$; Output: $y$, where $y= m \cdot x$
\\
\hline 
669 & $v_1=(1+2)\cdot x$ ; $v_2=(1-2^3)\cdot v_1$ ; $y=-v_1-2^5\cdot v_2$ \\
\hline 
2217 & $v_1=(1+2^4)\cdot x$ ; $v_2=(1+2)\cdot x$ ; $v_3=v_1+2^3 \cdot v_2$ ; \\
     & $y=2^7 \cdot v_1+v_3$ \\
\hline 
181  & $v_1=(1+2) \cdot x$ ; $v_2=2^3 \cdot x+v_1$ ; $y=2^6\cdot v_1-v_2$ \\
\hline 
3135 & $v_1=(1+2) \cdot x$ ; $v_2=(1-2^6) \cdot x$ ; $y=2^{10}\cdot v_1-v_2$ \\
\hline 
473  & $v_1=(1+ 2^2) \cdot x$ ; $v_2= x - 2^3 \cdot v_1$ ; $y=2^9 \cdot x+v_2$ \\
\hline 
437  & $v_1=(1+2^2) \cdot x$ ; $v_2=2^5 \cdot x-v_1$ ; $y=v_1+ 2^4 \cdot v_2$ \\
\hline 
2399 & $v_1=(1+2^2) \cdot x$ ; $v_2=x+ 2^5 \cdot v_1$ ; $y=2^9 \cdot v_1 - v_{2}$ \\
\hline 
8    & $y=2^3 \cdot x$ \\
\hline
\end{tabular}
\end{table}

Finally,
considering the above
quantities and
applying~(\ref{z4.representation}),
the sought fixed-point representations 
are fully recovered.
Hardware implementation of the multiplier circuits, 
required by the 
12-bit approximations above, 
is accomplished by using the method of 
Dempster and Macleod~\cite{dempster1994constant,gustafsson2006simplified}.
This method is known to be optimal for constant 
integer multiplier circuits. 

In this multiplierless method, 
the minimum number of 2-input adders are 
used for each constant integer multiplier. 
Wired shifts that perform ``costless'' multiplications by powers of two 
are used in each constant integer multiplier. 
Here, an enhancement to the Dempster-Macleod method is made for the constant 
integer multiplier circuits: the number of adder-bits is minimized, 
rather than the number of 2-input adders, 
yielding a smaller overall design.

Accordingly,
the multiplications by non powers of two 
shown in expressions (\ref{eq.frs.a})-(\ref{eq.frs.d})
can be algorithmically implemented as 
described in Table~\ref{table.fast.multiplication}.
Fig.~\ref{fig.frs.1.2} 
and~\ref{fig.frs.3.4}
depict the corresponding pipeline implementation.
Here, the various stages of the pipelined FRS architectures are shown
by having FIFO registers (consisting of parallel delay flip-flops (D-FFs)) vertically aligned in the figures.
Vertically aligned D-FFs indicate the same computation point
in a pipelined constant coefficient multiplication within the FRS.

\subsection{FRS based on expansion factor scaling}

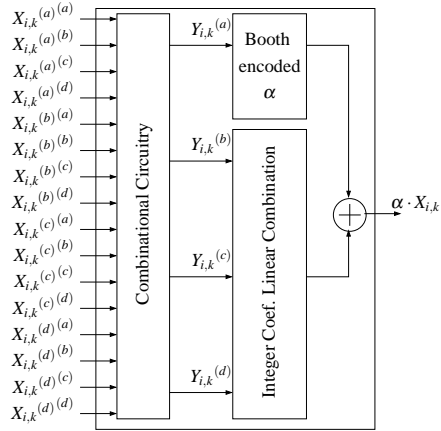
\begin{figure}
\begin{center}
  \input{rjdsc-fig-6.pstex_t}
\end{center}
\caption{Block diagram of the proposed AI decoding based on expansion factors.}
\label{fig_arch}
\end{figure}

The set of exact values given in~\eqref{exact-values-of-z}
suggests further relations among those quantities.
Indeed,
it may be established the following relations:
\begin{align*}
z_1^2&= 4 + z_1z_2,  &   z_2^2&=4-z_1z_2,
\\
z_1^2z_2&= 2\cdot(z_1+z_2), & z_1z_2^2&=2\cdot(z_1-z_2),
\\
z_1^2z_2^2&=8. & &
\end{align*}
These identities indicate that a new design
can be fostered.
In fact,
by substituting the above relations 
into~\eqref{eq-unfold-a}--\eqref{eq-unfold-d},
we have the following
expressions:
\begin{equation*}
\begin{split}
\getA{X_{i,k}} 
=& 
\getA{\getA{X_{i,k}}} 
+ 
\getB{\getA{X_{i,k}}} z_1 
+
\\
&
\getC{\getA{X_{i,k}}} z_2 
+
\getD{\getA{X_{i,k}}} z_1 z_2
,
\end{split}
\end{equation*}

\begin{equation*}
\begin{split}
\getB{X_{i,k}} z_1 
=& 
4 \cdot \getB{\getB{X_{i,k}}}
+
\left( 2 \cdot \getD{\getB{X_{i,k}}} + \getA{\getB{X_{i,k}}} \right)
z_1
+
\\
&
2 \cdot \getD{\getB{X_{i,k}}} z_2
+
\left( \getB{\getB{X_{i,k}}} + \getC{\getB{X_{i,k}}}  \right) z_1 z_2
,
\end{split}
\end{equation*}

\begin{equation*}
\begin{split}
\getC{X_{i,k}} z_2
=&
4 \cdot \getC{\getC{X_{i,k}}}
+
2 \cdot \getD{\getC{X_{i,k}}} z_1
+
\\
&
\left( \getA{\getC{X_{i,k}}} - 2\cdot \getD{\getC{X_{i,k}}} \right) z_2
+
\\
&
\left( \getB{\getC{X_{i,k}}} - \getC{\getC{X_{i,k}}}  \right) z_1 z_2
,
\end{split}
\end{equation*}

\begin{equation*}
\begin{split}
\getD{X_{i,k}} z_1 z_2
=&
8 \cdot \getD{\getD{X_{i,k}}}
+
2 \cdot \left( \getB{\getD{X_{i,k}}} + \getC{\getD{X_{i,k}}} \right) z_1
+
\\
&
2 \cdot \left( \getB{\getD{X_{i,k}}} - \getC{\getD{X_{i,k}}} \right) z_2
+
\getA{\getD{X_{i,k}}}
z_1 z_2
.
\end{split}
\end{equation*}
Notice that the output value
$X_{i,k}$ is the summation of the above quantities.
Therefore,
by grouping the terms on $\{ 1, z_1, z_2, z_1z_2 \}$,
we can express $X_{i,k}$
by the following summation:
\begin{align}
\label{eq-X-Y}
X_{i,k} 
=
\getA{Y_{i,k}}
+
\getB{Y_{i,k}}
z_1
+
\getC{Y_{i,k}}
z_2
+
\getD{Y_{i,k}}
z_1 z_2
,
\end{align}
where
\begin{equation}
\begin{split}
\getA{Y_{i,k}} =& 
\getA{\getA{X_{i,k}}} 
+ 
4\cdot \left(\getB{\getB{X_{i,k}}} + \getC{\getC{X_{i,k}}} \right)
+
\\
& 
8 \cdot \getD{\getD{X_{i,k}}},
\end{split}
\end{equation}

\begin{equation}
\begin{split}
\getB{Y_{i,k}} =&
\getB{\getA{X_{i,k}}} + \getA{\getB{X_{i,k}}} 
+
2 \cdot \left( \getD{\getB{X_{i,k}}}  +
\right. \\
& \left. 
\getD{\getC{X_{i,k}}} + \getB{\getD{X_{i,k}}} + \getC{\getD{X_{i,k}}} \right)
,
\end{split}
\end{equation}

\begin{equation}
\begin{split}
\getC{Y_{i,k}} =& 
\getC{\getA{X_{i,k}}} + \getA{\getC{X_{i,k}}} 
+
2 \cdot \left( \getD{\getB{X_{i,k}}} - 
\right.
\\
&
\left.
\getD{\getC{X_{i,k}}} + \getB{\getD{X_{i,k}}} - \getC{\getD{X_{i,k}}} \right)
,
\end{split}
\end{equation}

\begin{equation}
\begin{split}
\getD{Y_{i,k}} =& 
\getD{\getA{X_{i,k}}} + \getB{\getB{X_{i,k}}} + \getC{\getB{X_{i,k}}} + 
\\
&
\getB{\getC{X_{i,k}}} - \getC{\getC{X_{i,k}}} + \getA{\getD{X_{i,k}}}
.
\end{split}
\end{equation}
Quantities $\getQ{Y_{i,k}}$, $q\in\{a,b,c,d\}$,
require extremely simple arithmetic to be computed.
These operations are represented by the combinational block
in Fig.~\ref{fig_arch}.
We now turn to the problem of efficiently
evaluate~\eqref{eq-X-Y},
which depends on $z_1$, $z_2$, and $z_1z_2$.

A possibility is to employ an expansion factor
that could simultaneously
scale the quantities
$z_1$, $z_2$, and $z_1z_2$
into integer values.
This would facilitate the usage of integer arithmetic.
Such approach
has been often employed by 
integer transform designers~\cite{plonka2004global,cintra2011approximation}.
A good exposition on this method and related schemes 
is found in~\cite[Ch.~5]{britanak2007discrete}.

In mathematical terms,
we have the following problem.
Let the quantities
$z_1$, $z_2$, and $z_1z_2$
form a vector
$
\bm{\zeta} =
\begin{bmatrix}
z_1 & z_2 & z_1z_2
\end{bmatrix}^T
$.
An expansion factor~\cite[p.~274]{britanak2007discrete}
is the real number~$\alpha^\ast>1$
that satisfies the following minimization problem:
\begin{align}
\label{eq-min-prob}
\alpha^\ast
=
\arg
\min_{\alpha>1}
\Vert 
\alpha \cdot \bm{\zeta} - \operatorname{round}(\alpha \cdot \bm{\zeta})
\Vert
,
\end{align}
where $\Vert\cdot\Vert$ is a given error measure
and $\operatorname{round}(\cdot)$ is the rounding function.
We adopt the Euclidean norm as the error measure.
The presence of the rounding function
introduces several algebraic difficulties.
A closed-form solution for~\eqref{eq-min-prob}
is a non-trivial manipulation.
Thus,
we may resort to computational search.
Clearly,
additional restrictions must be imposed:
a limited search space and a given precision for $\alpha$.

In the range $\alpha \in [1, 256]$
with a precision of $10^{-4}$,
we could find the optimal value 
$\alpha^\ast = 167.2309$.
Thus,
we have the following scaling:
\begin{align*}
\alpha^\ast
\cdot
\begin{bmatrix}
z_1 \\ z_2 \\ z_1z_2
\end{bmatrix}
=
\begin{bmatrix}
436.995521744185\ldots \\
181.009471802748\ldots \\
473.00054429861\ldots
\end{bmatrix}
\approx
\begin{bmatrix}
437 \\
181 \\
473
\end{bmatrix}
.
\end{align*}
The error norm is approximately $10^{-2}$,
which is very low for this type of problem.

However,
notice that small values of $\alpha$ are desirable,
since they could scale $\bm{\zeta}$ into small integers,
which require a simple hardware design.
An analysis on the sub-optimal solutions for~\eqref{eq-min-prob}
shows that
$\alpha^\prime = 4.5961$ furnishes the following scaling:
\begin{align*}
\alpha^\prime
\cdot
\begin{bmatrix}
z_1 \\ z_2 \\ z_1z_2
\end{bmatrix}
=
\begin{bmatrix}
   12.01031370924931\ldots \\
    4.97483482672658\ldots \\
   12.99986988195626\ldots
\end{bmatrix}
\approx
\begin{bmatrix}
12 \\
5 \\
13
\end{bmatrix}
.
\end{align*}
In this case, 
the resulting integers are relatively small
and
the error norm is in the order of $10^{-2}$.

Now we are in position to address the computation of \eqref{eq-X-Y}.
Considering a given expansion factor $\alpha$,
we can write:
\begin{equation}
\label{eq2}
\begin{split}
X_{i,k}
=
\frac{1}{\alpha}
&
\left(
\alpha \cdot \getA{X_{i,k}}
+
m_1
\cdot
\getB{X_{i,k}}
+
\right.
\\
&
\left.
m_2
\cdot
\getC{X_{i,k}}
+
m_3
\cdot
\getD{X_{i,k}}
\right)
,
\end{split}
\end{equation}
where $m_1$, $m_2$, and $m_3$
are the integer constants implied by 
the expansion factor $\alpha$.
In particular,
these constants are 
$\{ 437, 181, 473 \}$,
for $\alpha = \alpha^\ast$,
and
$\{ 12, 5, 13 \}$,
for $\alpha = \alpha^\prime$.
Notice that~\eqref{eq2}
consists of a linear combination.

Because constants $m_1$, $m_2$, and $m_3$ are integers,
associate multiplications can be efficiently implemented in hardware.
Considering common subexpression elimination (CSE),
these multiplications are reduced to 
additions and shift operations,
requiring minimal amount of hardware resources.
For the set $\{ 437, 181, 473 \}$,
we have the following CSE manipulation:
\begin{align*}
437 \cdot \getB{Y_{i,k}} 
+
&
181 \cdot \getC{Y_{i,k}}
+ 
473 \cdot \getD{Y_{i,k}} 
=  
\\
&
473 \cdot \left(\getB{Y_{i,k}} + \getC{Y_{i,k}} + \getD{Y_{i,k}} \right)
-
\\
&
36 \cdot \left(\getB{Y_{i,k}} + \getC{Y_{i,k}} \right) 
-
\\
&
256 \cdot \getC{Y_{i,k}}
.
\end{align*}
This computation requires only eight additions.
Analogously,
for the set $\{ 12, 5, 13 \}$,
CSE yields:
\begin{align*}
12 \cdot \getB{Y_{i,k}} 
+
&
5 \cdot \getC{Y_{i,k}}
+ 
13 \cdot \getD{Y_{i,k}} 
=  
\\
&
8 \cdot \left(\getB{Y_{i,k}} + \getD{Y_{i,k}} \right)
+
\\
&
4 \cdot \left(\getB{Y_{i,k}} + \getC{Y_{i,k}} + \getD{Y_{i,k}} \right) 
+ 
\\
&
\getD{Y_{i,k}} + \getC{Y_{i,k}}
.
\end{align*}
Five additions are necessary.
Above calculations are represented by
the integer coefficient block in Fig.~\ref{fig_arch}.

The remaining multiplication in~\eqref{eq2}
is the one by $\alpha$,
which can be implemented according to the
Booth encoding representation.
Table~\ref{booth}
brings the required Booth encoding for
$\alpha^\ast = 167.2309$ and $\alpha^\prime = 4.5961$.

\begin{table}
\centering{}
\caption{Booth encoding of the expansion factors $\alpha$}
\label{booth}
\begin{tabular}{cc}
\hline 
$\alpha$ & Representation \\
\hline 
4.5961 & $2^2 + 2^{-1} + 2^{-4} + 2^{-5} + 2^{-9}$
\\
\hline 
167.2309 & $2^{7} + 2^{5} + 2^{3} - 2^{0} + 2^{-2} - 2^{-6} - 2^{-8}$
\\
\hline
\end{tabular}
\end{table}

The global multiplication by $1/\alpha$
is not problematic.
Indeed,
it can be embedded into
subsequent signal processing stages after the DCT operation.
Typically,
it is absorbed into the quantizer.
This approach has been employed in several
DCT 
architectures~\cite{bouguezel2008low,bouguezel2011parametric,cintra2011approximation}.

Fig.~\ref{fig_arch}
depicts the full block diagram of the discussed computing scheme.
Eight separate instances of this block
are necessary to compute
coefficients $X_{i,0}$ to $X_{i,7}$,
for each $i$.

\section{On-{FPGA} Test and Measurement}
\label{section.test}

\begin{table*}
\centering
\caption{Success rates of the DCT coefficient computation
for various fixed-point bus widths $L$ and tolerance levels}
\label{tab:success}
\begin{tabular}{|>{\centering}m{1.1cm}|>{\centering}m{1.1cm}|>{\centering}m{1.3cm}|>{\centering}m{0.75cm}|>{\centering}m{1cm}|>{\centering}m{1cm}|>{\centering}m{1cm}|>{\centering}m{1cm}|>{\centering}m{1cm}|>{\centering}m{1cm}|>{\centering}m{1cm}|}
\cline{5-11} 
\multicolumn{4}{>{\centering}m{0.75cm}|}{} & \multicolumn{7}{c|}{Percentage Tolerance}\tabularnewline
\cline{1-11} 
Design & \multicolumn{2}{>{\centering}m{3.15cm}|}{FRS Method} & $L$ & 10\% & 5\% & 1\% & 0.1\% & 0.05\% & 0.01\% & 0.005\%\tabularnewline
\hline
1 & \multicolumn{2}{>{\centering}m{3.5cm}|}{\multirow{2}{3.5cm}{\centering Dempster-Macleod}}& 4 & 99.9672 & 99.9203 & 99.6422 & 96.3563 & 92.7109 & 64.8406 & 42.1719\tabularnewline
\cline{1-1} 
\cline{4-11} 
2 & \multicolumn{2}{>{\centering}m{2.4cm}|}{} & 8 & 99.9719 & 99.9344 & 99.6047 & 96.3250 & 92.7031 & 64.7313 & 41.9016\tabularnewline
\hline
3 & \multirow{4}{1.1cm}{\centering Expansion factor} & \multirow{2}{1.3cm}{\centering $\{12, 5, 13\}$} & 4 & 99.1844 & 98.2944 & 91.6822 & 55.1811 & 45.0667 & 30.6922 & 22.8633\tabularnewline
\cline{1-1} 
\cline{4-11} 
4 & & & 8 & 99.1289 & 98.2944 & 91.4978 & 55.0900 & 45.0289 & 30.7122 & 22.8844\tabularnewline
\cline{1-1} 
\cline{3-11} 
5 & & \multirow{2}{1.3cm}{\centering $\{437, 181, 473 \}$} & 4 & 99.9900 & 99.9822 & 99.9178 & 99.1111 & 98.2000 & 91.0667 & 83.1244\tabularnewline
\cline{1-1} 
\cline{4-11} 
6 & & & 8 & 99.9589 & 99.9511 & 99.8733 & 99.0389 & 98.1278 & 90.9867 & 83.1767\tabularnewline
\hline
\end{tabular}
\end{table*}

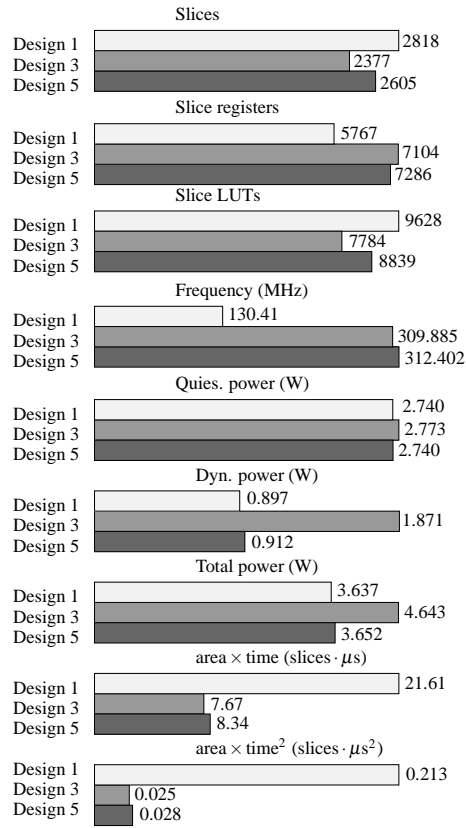
\begin{figure}
  \begin{center}
    \input{fpga_results_4.pstex_t}
  \end{center}
\caption{Resource utilization, speed of operation, and power consumption of the DCT designs given in Table~\ref{tab:success} on Xilinx Virtex-6 XC6VLX240T FPGA for input fixed-point wordlength $L=4$.}
\label{tab:util_speed1}
\end{figure}

\begin{figure}
  \begin{center}
    \input{fpga_results_8.pstex_t}
  \end{center}
\caption{Resource utilization, speed of operation, and power consumption of the DCT designs given in Table~\ref{tab:success} on Xilinx Virtex-6 XC6VLX240T FPGA for input fixed-point wordlength $L=8$.}
\label{tab:util_speed2}
\end{figure}

Six designs were implemented on Xilinx ML605 evaluation kit which is populated with a 
a Xilinx Virtex-6 XC6VLX240T device.
The designs included the three implementations of the 2D 8$\times$8 Arai AI DCT architecture with the two types 
of FRS described in Section~\ref{sec:frs} for fixed-point~4- and~8-bit wordlengths. Two versions of the expansion factor
FRSs are provided, corresponding to expansion factors $\alpha'=4.5941$ and $\alpha^*=167.2309$, resulting in 6 designs in total.
The proposed designs are listed in Table~\ref{tab:success}.

The JTAG interface was used to input the test 8$\times$8 \mbox{2-D} DCT arrays to the device from the \textsc{Matlab}
workspace.
Then
the measured outputs were returned to the \textsc{Matlab} workspace 
via the same interface. 
Hardware computed coefficients were compared to its numerical evaluation furnished by \textsc{Matlab} signal processing 
toolbox.

\subsection{On-chip Verification using Success Rates}

As a figure of merit, 
we considered the success rate 
defined as the percentage of coefficients 
which are within the error limit of $\pm e\,\%$. 
For $e = \{ 0.005, 0.01, 0.05, 0.1, 1, 5, 10 \}$,
the success rates
were measured as given in the Table~\ref{tab:success}.
Input wordlengths~$L$ was set to 4 or 8~bits.
The 8-bit size is the typical video processing configuration.
The proposed AI architectures enjoy overflow-free bit-growth
at each stage throughout the AI encoded structure thereby ensuring that
all sources of error are at the FRS and there only. 
Results show that
the FRS based on the expansion factor approach
for $\{437, 181, 473 \}$ 
(Designs 5 and 6) 
offers a significant improvement in accuracy
when compared to remaining FRS architectures.
 
\subsection{FPGA Resource Consumption}

The resource consumption of the proposed architectures
on Xilinx Virtex-6 XC6VLX240T device are shown in Fig.~\ref{tab:util_speed1}
for 
$L=4~\text{bits}$.
Fig.~\ref{tab:util_speed2} brings analogous information
for $L=8~\text{bits}$.
Here, 
FPGA resources are measured in terms of slices, 
slice registers,
and slice look-up-tables (LUTs).
Designs~3 and~4,
which use 
the FRS based on the expansion factor approach
for $\{ 12, 5, 13 \}$,
consumed the least resources in the device 
and has the worst accuracy of the three designs
(Table~\ref{tab:success}).
Moreover, even though Designs~5 and~6 
(FRS based on expansion factor approach for
$\{437,181,473\}$)
possesses superior accuracy 
when compared to 
Designs~1 and~2 
(FRS based on Dempster-Macleod method),
they 
consume less hardware resources.    
Overall
the FRS step of the proposed architectures
require a considerable amount of area
when compared to the 
AI steps of the architecture.

\subsection{Clock Speed, Block Rate, Frame Rate}

\begin{table}
  \caption{Frame rates and block rates achieved by the implemented designs for a video of resolution 1920$\times$1080}
  \centering
  \begin{tabular}{|>{\centering}p{.7cm}|>{\centering}p{1.2cm}|>{\centering}p{1.2cm}|>{\centering}p{1.3cm}|}
    \hline 
    Design & Freq. (MHz) & Block rate (MHz) & Frame rate (Hz)\tabularnewline
    \hline 
    1&130.410 & 16.30 & 503.08\tabularnewline
    \hline 
    2&123.120 & 15.39 & 475.00\tabularnewline
    \hline 
    3&309.855 & 38.73 & 1195.37\tabularnewline
    \hline 
    4&300.391 & 37.55 & 1158.95\tabularnewline
    \hline 
    5&312.402 & 39.05 & 1205.25\tabularnewline
    \hline
    6&307.787 & 38.47 & 1187.35\tabularnewline
    \hline
  \end{tabular}
  \label{tab:rates}
\end{table}

Frame rates and block rates achieved by the implemented designs for 
video at resolution 1920$\times$1080 is shown in Table~\ref{tab:rates}.
The design having the best throughput was Design~5, which operates on 4-bit inputs.
In Design~5, the maximum 8$\times$8 \mbox{2-D} DCT block rate is 39.05~MHz for a 312.402~MHz clock.
Assuming an input video resolution of 1920$\times$ 1080 pixels per frame, 
we obtained a real-time
computation of the \mbox{2-D} 8$\times$8 DCT
at 1205.25 frames per second.
In Design~6, 
we describe the common 8-bit input case, 
where the clock is now slightly reduced to~307.787~MHz, 
yielding an $8\times8$ block rate of~38.787~MHz, 
and a frame rate of~1187.35~frames per second for the same image size as above.
In all cases, if the 2-D DCT core is
eventually embedded in a real-time video processor, 
the pixel rate is eight fold the clock frequency of the DCT core 
(due to the downsampling by eight in the signal flow graph).
For example, 
a potential pixel rate of $\approx$2.499~GHz and $\approx$2.462~GHz, 
for Designs~5 and~6, 
may be possible.

\subsection {Xilinx Power Consumption and Critical Path}

The total power consumption of FPGA circuits consist of 
the sum of dynamic and quiescent power consumptions. 
Both estimated dynamic and quiescent power consumptions obtained 
from the design tools for 
the Xilinx Virtex-6 XC6VLX240T device are provided 
in Fig.~\ref{tab:util_speed1} and Fig.~\ref{tab:util_speed2}.

\subsection{Area-Time Complexity Metrics}

Estimates for VLSI area-time complexity metrics are provided for all designs
are given in Fig.~\ref{tab:util_speed1} ($L=4$)
and 
Fig.~\ref{tab:util_speed2} ($L=8$), 
respectively.
In general, 
the area-time metric measures complexity of VLSI circuits where chip real-estate is important over speed, while metric area-time$^2$ is used
often for VLSI circuits where speed is of paramount concern. 
We provide both metrics to offer a broad overview of the area-time complexity
levels present in the proposed architectures as 
a function of input size and choice of FRS algorithm. 

The architectures are free of general purpose multipliers.

\subsection{Overall Comparison with Existing Architectures}

Fixed point 
VLSI implementations that are directly comparable to 
the proposed architecture are compared 
in detail 
in Table~\ref{tab:fixed_prev}.
Table~\ref{tab:ai_prev}
brings comparisons
to AI-based architectures.
For brevity and without loss of generality, we chose designs 2 and 6 for the purpose of comparison. These are 
aimed at 8-bit input signals and are examples of the Dempster-Macleod and expansion factor FRS
algorithms. A synopsis of both fixed-point and AI-based 2D-DCT
circuits under comparison in Tables~~\ref{tab:fixed_prev} and  ~\ref{tab:ai_prev} was provided in Section~\ref{section.algebraic}.

\begin{table*}
  \centering
 \begin{threeparttable}[b]
  \caption{Comparison of the proposed implementation with  published fixed point implementations\label{tab:fixed_prev}}
  \begin{tabular}{|>{\centering}m{1.7cm}|>{\centering}m{1.2cm}|>{\centering}m{1.2cm}|>{\centering}m{1.2cm}|>{\centering}m{1.2cm}|>{\centering}m{1.2cm}|>{\centering}m{1.2cm}|>{\centering}m{1.2cm}|>{\centering}>{\bfseries}m{1.4cm}|>{\centering}>{\bfseries}m{1.4cm}|}
    \cline{2-10} 
    \multicolumn{1}{>{\centering}m{1.2cm}|}{} & \multirow{2}{1.2cm}{Lin~et~al.\\\ \ \ \cite{lin_yu_van}} & \multirow{2}{1.2cm}{Shams~et~al.\\\ \ \ \cite{shams_pan_bayoumi}} & \multirow{2}{1.2cm}{Madisetti\\et~al.~\cite{madisetti_willson}} & \multirow{2}{1.2cm}{Guo~et~al.\\\ \ \ \ \cite{guo_ju}} & \multirow{2}{1.2cm}{Tumeo~et~al.\\\ \ \ \cite{tumeo}} & \multirow{2}{1.2cm}{Sun~et~al.\\\ \ \ \cite{sun_donner}} & \multirow{2}{1.2cm}{Chen~et~al.\\\ \ \ \cite{chen_chang}} & \multicolumn{2}{>{\centering}m{3.2cm}|}{Proposed architectures} \tabularnewline
    \cline{9-10} 
    \multicolumn{1}{>{\centering}m{1.2cm}|}{}& & & & & & & & Design 2 & Design 6 \tabularnewline
    \hline 
    Measured results & No & No & No & No & No & No & No & Yes & Yes\tabularnewline
    \hline 
    Structure & Single \mbox{2-D} DCT & Two \mbox{1-D} DCT +TMEM\tnote{$\dagger$} & Single \mbox{1-D} DCT +TMEM\tnote{$\dagger$} & Single \mbox{1-D} DCT +TMEM\tnote{$\dagger$} & Single \mbox{1-D} DCT +TMEM\tnote{$\dagger$} & Two \mbox{1-D} DCT +TMEM\tnote{$\dagger$} & Single \mbox{1-D} DCT +TMEM\tnote{$\dagger$} & See Fig.~\ref{fig1} & See Fig.~\ref{fig1}\tabularnewline
    \hline 
    Multipliers & 1 & 0 & 7 & 0 & 4 & 0 & 0 & 0 & 0\tabularnewline
    \hline 
    Operating frequency (MHz)& 100  & N/A & 100 & 110 & 107 & 149 & 167 & 123.12 & 307.79\tabularnewline
    \hline
    8$\times$8~Block~rate\\$\times10^6\mathrm{s}^{-1}$ & 1.5625 & N/A & 1.562 & 3.4375 & 1.3375 & 2.328 & 2.609 & 15.39\tnote{$\ast$} & 38.625\tnote{$\ast$}\tabularnewline
    \hline
    Pixel rate\\$\times10^6\mathrm{s}^{-1}$ & 100 & N/A & 100 & 220 & 85.6 & 149 & 167 & 984.96\tnote{$\ddagger$} & 2462.32\tnote{$\ddagger$}\tabularnewline
    \hline
    Implementation technology & 0.13$\mu$m CMOS & N/A & 0.8$\mu$m CMOS & 0.35$\mu$m CMOS & Xilinx XC2VP30& 
    Xilinx XC2VP30 & 0.18$\mu$m CMOS & \scriptsize Xilinx XC6VLX240T & \scriptsize Xilinx XC6VLX240T \tabularnewline
    \hline
    Coupled quantization noise & Yes & Yes & Yes & Yes & Yes & Yes & Yes & No & No\tabularnewline
    \hline
    Independently adjustable precision & No & No & No & No & No & No & No & Yes & Yes\tabularnewline
    \hline
  \end{tabular}
  \begin{tablenotes}
  \item [$\dagger$] Row column transpose buffer.~~\item [$\ast$] Block rate$ = F_{clock}/8$.~~\item [$\ddagger$] Pixel rate $ = F_s$. 
  \end{tablenotes}
\end{threeparttable}
\end{table*}

\begin{table*}
  \centering
  \begin{threeparttable}[b]
  \caption{Comparison of the proposed implementation with published algebraic integer implementations\label{tab:ai_prev}}
  \begin{tabular}{|>{\centering}m{2cm}|>{\centering}m{1.8cm}|>{\centering}m{1.8cm}|>{\centering}m{1.8cm}|>{\centering}>{\bfseries}m{1.8cm}|>{\centering}>{\bfseries}m{1.8cm}|}
    \cline{2-6} 
    \multicolumn{1}{>{\centering}m{2cm}|}{} & \multirow{2}{1.3cm}{Nandi~et~al.\\\ \ \ \ \cite{nandi}} & \multirow{2}{1.3cm}{Jullien~et~al.\\\ \ \ \ \cite{fu_2004}} & \multirow{2}{1.3cm}{Wahid~et~al.\\\ \ \ \ \cite{wahid_murtuza}} & \multicolumn{2}{>{\centering}m{4cm}|}{Proposed architectures} \tabularnewline
    \cline{5-6} 
    \multicolumn{1}{>{\centering}m{2cm}|}{}& & & & Design 2 & Design 6 \tabularnewline
    \hline 
    Measured\\results & No & No & No & Yes & Yes\tabularnewline
    \hline 
    Structure & Single \mbox{1-D} DCT +Mem. bank & Two \mbox{1-D} DCT +Dual port RAM & Two \mbox{1-D} DCT +TMEM\tnote{$\dagger$} & See Fig.~\ref{fig1} & See Fig.~\ref{fig1}\tabularnewline
    \hline 
    Multipliers & 0 & 0 & 0 & 0 & 0\tabularnewline
    \hline 
    Exact 2D AI computation & No & No & No & Yes & Yes\tabularnewline
    \hline 
    Operating\\frequency\\(MHz) & N/A & 75 & 194.7 & 123.12 & 307.79\tabularnewline
    \hline
    8$\times$8 Block rate $\times10^6\mathrm{s}^{-1}$ & 7.8125 & 1.171 & 3.042 & 15.39\tnote{$\ast$} & 38.625\tnote{$\ast$}\tabularnewline
    \hline
    Pixel rate\\$\times10^6\mathrm{s}^{-1}$ & 125 & 75 & 194.7 & 984.96\tnote{$\ddagger$} & 2462.32\tnote{$\ddagger$}\tabularnewline
    \hline
    Implementation technology & Xilinx XC5VLX30 & 0.18$\mu$m\\CMOS & 0.18$\mu$m\\CMOS & Xilinx XC6VLX240T & Xilinx XC6VLX240T\tabularnewline
    \hline
    Coupled quantization\\noise & Yes & Yes & Yes & No & No\tabularnewline
    \hline
    Independently adjustable\\precision & No  & No & No & Yes & Yes \tabularnewline
    \hline
    FRS between row-column stages & No & Yes & Yes & No & No \tabularnewline
    \hline
  \end{tabular}
  \begin{tablenotes}
  \item [$\dagger$] Row column transpose buffer.~~\item [$\ast$] Block rate$ = F_{clock}/8$.~~\item [$\ddagger$] Pixel rate $ = F_s$. 
  \end{tablenotes}
\end{threeparttable}
\end{table*}

\section{Conclusions}

\label{section.conclusions}

A time-multiplexed systolic-array hardware architecture
is proposed
for the real-time computation of the bivariate AI encoded \mbox{2-D} Arai DCT.
The architecture is the first \mbox{2-D} AI encoded DCT hardware
that operates completely in the AI domain.
This not only makes the proposed system completely multiplier-free,
but also quantization free up to the final output channels.

Our architecture employs a novel AI-TB,
which facilitates real-time data transposition.
The \mbox{2-D} separable DCT operation is entirely performed
in the AI domain.
Indeed,
the architecture does not have intermediate FRS sections
between the column- and row-wise AI-based Arai DCT operations.
This
makes
the quantization noise only appear at
the final output stage of the architecture:
the
single FRS section.

The location of the FRS at the final output stage
results
in the complete decoupling of quantization noise between
the 64~parallel coefficient channels of the \mbox{2-D} DCT.
This fact is noteworthy because it enables
the independent selection of precision
for each of the 64~channels
without having any effect on the
speed, power, complexity, or noise level of the remaining channels.

Two algorithms for the FRS are proposed, 
numerically optimized, 
analyzed, 
hardware implemented,
and tested 
with the proposed \mbox{2-D} AI encoded section.
The architectures are physically implemented
for input precision of 4 and 8 bits, and fully verified on-chip.
Of particular relevance is the commonly required 8-bit realization, which is operational
at a clock frequency of 307.787~MHz
on a Xilinx Virtex-6 XC6VLX240T FPGA device (see~Design~6). 
This implies a
$8\times8$ block rate of 38.47~MHz and 
a \emph{potential} pixel rate of
$\approx$2.462~GHz \emph{if} the proposed \mbox{2-D} DCT core is embedded in a real-time video processing system.
The frame rate for standard HD video at $1920\times1080$ resolution
is 
$\approx$1187.35~Hz 
assuming 8-bit input words and core clock frequency of~307.787~MHz.

The proposed architecture achieves 
complete elimination of quantization noise
coupling between DCT coefficients, 
which is present in published \mbox{2-D} DCT architectures based on
both fixed-point arithmetic as well as row-column 
8-point
Arai DCT cores that have FRS sections between row- and column-wise transforms. 
The proposed designs allows each of the 64~coefficients to be computed at 
64~different precision levels, where each choice of precision only affects that particular coefficient. 
This allows full control of the \mbox{2-D} DCT computation to any
degree of precision desired by the designer.

\section*{Acknowledgments}

This work was partially supported by CNPq and FACEPE.

{\small
\bibliographystyle{IEEEtran}
\bibliography{ref1}
}

\end{document}

%% file: rjdsc-AI-TB.pstex_t
\begin{picture}(0,0)%
\includegraphics{rjdsc-AI-TB.pstex}%
\end{picture}%
\setlength{\unitlength}{3522sp}%
\begingroup\makeatletter\ifx\SetFigFont\undefined%
\gdef\SetFigFont#1#2#3#4#5{%
  \reset@font\fontsize{#1}{#2pt}%
  \fontfamily{#3}\fontseries{#4}\fontshape{#5}%
  \selectfont}%
\fi\endgroup%
\begin{picture}(3582,5154)(349,-5203)
\put(1576,-286){\makebox(0,0)[b]{\smash{{\SetFigFont{5}{6.0}{\rmdefault}{\mddefault}{\updefault}{\color[rgb]{0,0,0}$z^{-1}$}%
}}}}
\put(1891,-286){\makebox(0,0)[b]{\smash{{\SetFigFont{5}{6.0}{\rmdefault}{\mddefault}{\updefault}{\color[rgb]{0,0,0}$z^{-1}$}%
}}}}
\put(2206,-286){\makebox(0,0)[b]{\smash{{\SetFigFont{5}{6.0}{\rmdefault}{\mddefault}{\updefault}{\color[rgb]{0,0,0}$z^{-1}$}%
}}}}
\put(2521,-286){\makebox(0,0)[b]{\smash{{\SetFigFont{5}{6.0}{\rmdefault}{\mddefault}{\updefault}{\color[rgb]{0,0,0}$z^{-1}$}%
}}}}
\put(2836,-286){\makebox(0,0)[b]{\smash{{\SetFigFont{5}{6.0}{\rmdefault}{\mddefault}{\updefault}{\color[rgb]{0,0,0}$z^{-1}$}%
}}}}
\put(3151,-286){\makebox(0,0)[b]{\smash{{\SetFigFont{5}{6.0}{\rmdefault}{\mddefault}{\updefault}{\color[rgb]{0,0,0}$z^{-1}$}%
}}}}
\put(3466,-286){\makebox(0,0)[b]{\smash{{\SetFigFont{5}{6.0}{\rmdefault}{\mddefault}{\updefault}{\color[rgb]{0,0,0}$z^{-1}$}%
}}}}
\put(2206,-841){\makebox(0,0)[b]{\smash{{\SetFigFont{5}{6.0}{\rmdefault}{\mddefault}{\updefault}{\color[rgb]{0,0,0}$z^{-1}$}%
}}}}
\put(2521,-841){\makebox(0,0)[b]{\smash{{\SetFigFont{5}{6.0}{\rmdefault}{\mddefault}{\updefault}{\color[rgb]{0,0,0}$z^{-1}$}%
}}}}
\put(2836,-841){\makebox(0,0)[b]{\smash{{\SetFigFont{5}{6.0}{\rmdefault}{\mddefault}{\updefault}{\color[rgb]{0,0,0}$z^{-1}$}%
}}}}
\put(3151,-841){\makebox(0,0)[b]{\smash{{\SetFigFont{5}{6.0}{\rmdefault}{\mddefault}{\updefault}{\color[rgb]{0,0,0}$z^{-1}$}%
}}}}
\put(3466,-841){\makebox(0,0)[b]{\smash{{\SetFigFont{5}{6.0}{\rmdefault}{\mddefault}{\updefault}{\color[rgb]{0,0,0}$z^{-1}$}%
}}}}
\put(1882,-837){\makebox(0,0)[b]{\smash{{\SetFigFont{5}{6.0}{\rmdefault}{\mddefault}{\updefault}{\color[rgb]{0,0,0}$z^{-1}$}%
}}}}
\put(1576,-837){\makebox(0,0)[b]{\smash{{\SetFigFont{5}{6.0}{\rmdefault}{\mddefault}{\updefault}{\color[rgb]{0,0,0}$z^{-1}$}%
}}}}
\put(2206,-1411){\makebox(0,0)[b]{\smash{{\SetFigFont{5}{6.0}{\rmdefault}{\mddefault}{\updefault}{\color[rgb]{0,0,0}$z^{-1}$}%
}}}}
\put(2521,-1411){\makebox(0,0)[b]{\smash{{\SetFigFont{5}{6.0}{\rmdefault}{\mddefault}{\updefault}{\color[rgb]{0,0,0}$z^{-1}$}%
}}}}
\put(2836,-1411){\makebox(0,0)[b]{\smash{{\SetFigFont{5}{6.0}{\rmdefault}{\mddefault}{\updefault}{\color[rgb]{0,0,0}$z^{-1}$}%
}}}}
\put(3151,-1411){\makebox(0,0)[b]{\smash{{\SetFigFont{5}{6.0}{\rmdefault}{\mddefault}{\updefault}{\color[rgb]{0,0,0}$z^{-1}$}%
}}}}
\put(3466,-1411){\makebox(0,0)[b]{\smash{{\SetFigFont{5}{6.0}{\rmdefault}{\mddefault}{\updefault}{\color[rgb]{0,0,0}$z^{-1}$}%
}}}}
\put(1882,-1407){\makebox(0,0)[b]{\smash{{\SetFigFont{5}{6.0}{\rmdefault}{\mddefault}{\updefault}{\color[rgb]{0,0,0}$z^{-1}$}%
}}}}
\put(1576,-1407){\makebox(0,0)[b]{\smash{{\SetFigFont{5}{6.0}{\rmdefault}{\mddefault}{\updefault}{\color[rgb]{0,0,0}$z^{-1}$}%
}}}}
\put(1261,-1401){\makebox(0,0)[b]{\smash{{\SetFigFont{5}{6.0}{\rmdefault}{\mddefault}{\updefault}{\color[rgb]{0,0,0}$z^{-2}$}%
}}}}
\put(2206,-1996){\makebox(0,0)[b]{\smash{{\SetFigFont{5}{6.0}{\rmdefault}{\mddefault}{\updefault}{\color[rgb]{0,0,0}$z^{-1}$}%
}}}}
\put(2521,-1996){\makebox(0,0)[b]{\smash{{\SetFigFont{5}{6.0}{\rmdefault}{\mddefault}{\updefault}{\color[rgb]{0,0,0}$z^{-1}$}%
}}}}
\put(2836,-1996){\makebox(0,0)[b]{\smash{{\SetFigFont{5}{6.0}{\rmdefault}{\mddefault}{\updefault}{\color[rgb]{0,0,0}$z^{-1}$}%
}}}}
\put(3151,-1996){\makebox(0,0)[b]{\smash{{\SetFigFont{5}{6.0}{\rmdefault}{\mddefault}{\updefault}{\color[rgb]{0,0,0}$z^{-1}$}%
}}}}
\put(3466,-1996){\makebox(0,0)[b]{\smash{{\SetFigFont{5}{6.0}{\rmdefault}{\mddefault}{\updefault}{\color[rgb]{0,0,0}$z^{-1}$}%
}}}}
\put(1882,-1992){\makebox(0,0)[b]{\smash{{\SetFigFont{5}{6.0}{\rmdefault}{\mddefault}{\updefault}{\color[rgb]{0,0,0}$z^{-1}$}%
}}}}
\put(1576,-1992){\makebox(0,0)[b]{\smash{{\SetFigFont{5}{6.0}{\rmdefault}{\mddefault}{\updefault}{\color[rgb]{0,0,0}$z^{-1}$}%
}}}}
\put(1261,-1986){\makebox(0,0)[b]{\smash{{\SetFigFont{5}{6.0}{\rmdefault}{\mddefault}{\updefault}{\color[rgb]{0,0,0}$z^{-3}$}%
}}}}
\put(2206,-2581){\makebox(0,0)[b]{\smash{{\SetFigFont{5}{6.0}{\rmdefault}{\mddefault}{\updefault}{\color[rgb]{0,0,0}$z^{-1}$}%
}}}}
\put(2521,-2581){\makebox(0,0)[b]{\smash{{\SetFigFont{5}{6.0}{\rmdefault}{\mddefault}{\updefault}{\color[rgb]{0,0,0}$z^{-1}$}%
}}}}
\put(2836,-2581){\makebox(0,0)[b]{\smash{{\SetFigFont{5}{6.0}{\rmdefault}{\mddefault}{\updefault}{\color[rgb]{0,0,0}$z^{-1}$}%
}}}}
\put(3151,-2581){\makebox(0,0)[b]{\smash{{\SetFigFont{5}{6.0}{\rmdefault}{\mddefault}{\updefault}{\color[rgb]{0,0,0}$z^{-1}$}%
}}}}
\put(3466,-2581){\makebox(0,0)[b]{\smash{{\SetFigFont{5}{6.0}{\rmdefault}{\mddefault}{\updefault}{\color[rgb]{0,0,0}$z^{-1}$}%
}}}}
\put(1882,-2577){\makebox(0,0)[b]{\smash{{\SetFigFont{5}{6.0}{\rmdefault}{\mddefault}{\updefault}{\color[rgb]{0,0,0}$z^{-1}$}%
}}}}
\put(1576,-2577){\makebox(0,0)[b]{\smash{{\SetFigFont{5}{6.0}{\rmdefault}{\mddefault}{\updefault}{\color[rgb]{0,0,0}$z^{-1}$}%
}}}}
\put(1261,-2571){\makebox(0,0)[b]{\smash{{\SetFigFont{5}{6.0}{\rmdefault}{\mddefault}{\updefault}{\color[rgb]{0,0,0}$z^{-4}$}%
}}}}
\put(2206,-3166){\makebox(0,0)[b]{\smash{{\SetFigFont{5}{6.0}{\rmdefault}{\mddefault}{\updefault}{\color[rgb]{0,0,0}$z^{-1}$}%
}}}}
\put(2521,-3166){\makebox(0,0)[b]{\smash{{\SetFigFont{5}{6.0}{\rmdefault}{\mddefault}{\updefault}{\color[rgb]{0,0,0}$z^{-1}$}%
}}}}
\put(2836,-3166){\makebox(0,0)[b]{\smash{{\SetFigFont{5}{6.0}{\rmdefault}{\mddefault}{\updefault}{\color[rgb]{0,0,0}$z^{-1}$}%
}}}}
\put(3151,-3166){\makebox(0,0)[b]{\smash{{\SetFigFont{5}{6.0}{\rmdefault}{\mddefault}{\updefault}{\color[rgb]{0,0,0}$z^{-1}$}%
}}}}
\put(3466,-3166){\makebox(0,0)[b]{\smash{{\SetFigFont{5}{6.0}{\rmdefault}{\mddefault}{\updefault}{\color[rgb]{0,0,0}$z^{-1}$}%
}}}}
\put(1882,-3162){\makebox(0,0)[b]{\smash{{\SetFigFont{5}{6.0}{\rmdefault}{\mddefault}{\updefault}{\color[rgb]{0,0,0}$z^{-1}$}%
}}}}
\put(1576,-3162){\makebox(0,0)[b]{\smash{{\SetFigFont{5}{6.0}{\rmdefault}{\mddefault}{\updefault}{\color[rgb]{0,0,0}$z^{-1}$}%
}}}}
\put(1261,-3156){\makebox(0,0)[b]{\smash{{\SetFigFont{5}{6.0}{\rmdefault}{\mddefault}{\updefault}{\color[rgb]{0,0,0}$z^{-5}$}%
}}}}
\put(2206,-3751){\makebox(0,0)[b]{\smash{{\SetFigFont{5}{6.0}{\rmdefault}{\mddefault}{\updefault}{\color[rgb]{0,0,0}$z^{-1}$}%
}}}}
\put(2521,-3751){\makebox(0,0)[b]{\smash{{\SetFigFont{5}{6.0}{\rmdefault}{\mddefault}{\updefault}{\color[rgb]{0,0,0}$z^{-1}$}%
}}}}
\put(2836,-3751){\makebox(0,0)[b]{\smash{{\SetFigFont{5}{6.0}{\rmdefault}{\mddefault}{\updefault}{\color[rgb]{0,0,0}$z^{-1}$}%
}}}}
\put(3151,-3751){\makebox(0,0)[b]{\smash{{\SetFigFont{5}{6.0}{\rmdefault}{\mddefault}{\updefault}{\color[rgb]{0,0,0}$z^{-1}$}%
}}}}
\put(3466,-3751){\makebox(0,0)[b]{\smash{{\SetFigFont{5}{6.0}{\rmdefault}{\mddefault}{\updefault}{\color[rgb]{0,0,0}$z^{-1}$}%
}}}}
\put(1882,-3747){\makebox(0,0)[b]{\smash{{\SetFigFont{5}{6.0}{\rmdefault}{\mddefault}{\updefault}{\color[rgb]{0,0,0}$z^{-1}$}%
}}}}
\put(1576,-3747){\makebox(0,0)[b]{\smash{{\SetFigFont{5}{6.0}{\rmdefault}{\mddefault}{\updefault}{\color[rgb]{0,0,0}$z^{-1}$}%
}}}}
\put(1261,-3741){\makebox(0,0)[b]{\smash{{\SetFigFont{5}{6.0}{\rmdefault}{\mddefault}{\updefault}{\color[rgb]{0,0,0}$z^{-6}$}%
}}}}
\put(2206,-4336){\makebox(0,0)[b]{\smash{{\SetFigFont{5}{6.0}{\rmdefault}{\mddefault}{\updefault}{\color[rgb]{0,0,0}$z^{-1}$}%
}}}}
\put(2521,-4336){\makebox(0,0)[b]{\smash{{\SetFigFont{5}{6.0}{\rmdefault}{\mddefault}{\updefault}{\color[rgb]{0,0,0}$z^{-1}$}%
}}}}
\put(2836,-4336){\makebox(0,0)[b]{\smash{{\SetFigFont{5}{6.0}{\rmdefault}{\mddefault}{\updefault}{\color[rgb]{0,0,0}$z^{-1}$}%
}}}}
\put(3151,-4336){\makebox(0,0)[b]{\smash{{\SetFigFont{5}{6.0}{\rmdefault}{\mddefault}{\updefault}{\color[rgb]{0,0,0}$z^{-1}$}%
}}}}
\put(3466,-4336){\makebox(0,0)[b]{\smash{{\SetFigFont{5}{6.0}{\rmdefault}{\mddefault}{\updefault}{\color[rgb]{0,0,0}$z^{-1}$}%
}}}}
\put(1882,-4332){\makebox(0,0)[b]{\smash{{\SetFigFont{5}{6.0}{\rmdefault}{\mddefault}{\updefault}{\color[rgb]{0,0,0}$z^{-1}$}%
}}}}
\put(1576,-4332){\makebox(0,0)[b]{\smash{{\SetFigFont{5}{6.0}{\rmdefault}{\mddefault}{\updefault}{\color[rgb]{0,0,0}$z^{-1}$}%
}}}}
\put(1261,-4326){\makebox(0,0)[b]{\smash{{\SetFigFont{5}{6.0}{\rmdefault}{\mddefault}{\updefault}{\color[rgb]{0,0,0}$z^{-7}$}%
}}}}
\put(1261,-831){\makebox(0,0)[b]{\smash{{\SetFigFont{5}{6.0}{\rmdefault}{\mddefault}{\updefault}{\color[rgb]{0,0,0}$z^{-1}$}%
}}}}
\put(1036,-196){\makebox(0,0)[rb]{\smash{{\SetFigFont{7}{8.4}{\rmdefault}{\mddefault}{\updefault}{\color[rgb]{0,0,0}$\getA{X_{0,k}}$}%
}}}}
\put(1036,-736){\makebox(0,0)[rb]{\smash{{\SetFigFont{7}{8.4}{\rmdefault}{\mddefault}{\updefault}{\color[rgb]{0,0,0}$\getA{X_{1,k}}$}%
}}}}
\put(1036,-1306){\makebox(0,0)[rb]{\smash{{\SetFigFont{7}{8.4}{\rmdefault}{\mddefault}{\updefault}{\color[rgb]{0,0,0}$\getA{X_{2,k}}$}%
}}}}
\put(1036,-1891){\makebox(0,0)[rb]{\smash{{\SetFigFont{7}{8.4}{\rmdefault}{\mddefault}{\updefault}{\color[rgb]{0,0,0}$\getA{X_{3,k}}$}%
}}}}
\put(1036,-2476){\makebox(0,0)[rb]{\smash{{\SetFigFont{7}{8.4}{\rmdefault}{\mddefault}{\updefault}{\color[rgb]{0,0,0}$\getA{X_{4,k}}$}%
}}}}
\put(1036,-3061){\makebox(0,0)[rb]{\smash{{\SetFigFont{7}{8.4}{\rmdefault}{\mddefault}{\updefault}{\color[rgb]{0,0,0}$\getA{X_{5,k}}$}%
}}}}
\put(1036,-3646){\makebox(0,0)[rb]{\smash{{\SetFigFont{7}{8.4}{\rmdefault}{\mddefault}{\updefault}{\color[rgb]{0,0,0}$\getA{X_{6,k}}$}%
}}}}
\put(1036,-4231){\makebox(0,0)[rb]{\smash{{\SetFigFont{7}{8.4}{\rmdefault}{\mddefault}{\updefault}{\color[rgb]{0,0,0}$\getA{X_{7,k}}$}%
}}}}
\put(3781,-4741){\makebox(0,0)[lb]{\smash{{\SetFigFont{7}{8.4}{\rmdefault}{\mddefault}{\updefault}{\color[rgb]{0,0,0}$\getA{X_{7,k}}$}%
}}}}
\put(3781,-4336){\makebox(0,0)[lb]{\smash{{\SetFigFont{7}{8.4}{\rmdefault}{\mddefault}{\updefault}{\color[rgb]{0,0,0}$\getA{X_{7,k-7}}$}%
}}}}
\put(3781,-4156){\makebox(0,0)[lb]{\smash{{\SetFigFont{7}{8.4}{\rmdefault}{\mddefault}{\updefault}{\color[rgb]{0,0,0}$\getA{X_{6,k}}$}%
}}}}
\put(3781,-3751){\makebox(0,0)[lb]{\smash{{\SetFigFont{7}{8.4}{\rmdefault}{\mddefault}{\updefault}{\color[rgb]{0,0,0}$\getA{X_{6,k-7}}$}%
}}}}
\put(3781,-3571){\makebox(0,0)[lb]{\smash{{\SetFigFont{7}{8.4}{\rmdefault}{\mddefault}{\updefault}{\color[rgb]{0,0,0}$\getA{X_{5,k}}$}%
}}}}
\put(3781,-3166){\makebox(0,0)[lb]{\smash{{\SetFigFont{7}{8.4}{\rmdefault}{\mddefault}{\updefault}{\color[rgb]{0,0,0}$\getA{X_{5,k-7}}$}%
}}}}
\put(3781,-2986){\makebox(0,0)[lb]{\smash{{\SetFigFont{7}{8.4}{\rmdefault}{\mddefault}{\updefault}{\color[rgb]{0,0,0}$\getA{X_{4,k}}$}%
}}}}
\put(3781,-2581){\makebox(0,0)[lb]{\smash{{\SetFigFont{7}{8.4}{\rmdefault}{\mddefault}{\updefault}{\color[rgb]{0,0,0}$\getA{X_{4,k-7}}$}%
}}}}
\put(3781,-2401){\makebox(0,0)[lb]{\smash{{\SetFigFont{7}{8.4}{\rmdefault}{\mddefault}{\updefault}{\color[rgb]{0,0,0}$\getA{X_{3,k}}$}%
}}}}
\put(3781,-1996){\makebox(0,0)[lb]{\smash{{\SetFigFont{7}{8.4}{\rmdefault}{\mddefault}{\updefault}{\color[rgb]{0,0,0}$\getA{X_{3,k-7}}$}%
}}}}
\put(3781,-1816){\makebox(0,0)[lb]{\smash{{\SetFigFont{7}{8.4}{\rmdefault}{\mddefault}{\updefault}{\color[rgb]{0,0,0}$\getA{X_{2,k}}$}%
}}}}
\put(3781,-1411){\makebox(0,0)[lb]{\smash{{\SetFigFont{7}{8.4}{\rmdefault}{\mddefault}{\updefault}{\color[rgb]{0,0,0}$\getA{X_{2,k-7}}$}%
}}}}
\put(3781,-1231){\makebox(0,0)[lb]{\smash{{\SetFigFont{7}{8.4}{\rmdefault}{\mddefault}{\updefault}{\color[rgb]{0,0,0}$\getA{X_{1,k}}$}%
}}}}
\put(3781,-286){\makebox(0,0)[lb]{\smash{{\SetFigFont{7}{8.4}{\rmdefault}{\mddefault}{\updefault}{\color[rgb]{0,0,0}$\getA{X_{0,k-7}}$}%
}}}}
\put(3781,-691){\makebox(0,0)[lb]{\smash{{\SetFigFont{7}{8.4}{\rmdefault}{\mddefault}{\updefault}{\color[rgb]{0,0,0}$\getA{X_{0,k}}$}%
}}}}
\put(3781,-826){\makebox(0,0)[lb]{\smash{{\SetFigFont{7}{8.4}{\rmdefault}{\mddefault}{\updefault}{\color[rgb]{0,0,0}$\getA{X_{1,k-7}}$}%
}}}}
\put(3916,-3391){\makebox(0,0)[lb]{\smash{{\SetFigFont{5}{6.0}{\rmdefault}{\mddefault}{\updefault}{\color[rgb]{0,0,0}$\vdots$}%
}}}}
\put(3916,-1051){\makebox(0,0)[lb]{\smash{{\SetFigFont{5}{6.0}{\rmdefault}{\mddefault}{\updefault}{\color[rgb]{0,0,0}$\vdots$}%
}}}}
\put(3916,-511){\makebox(0,0)[lb]{\smash{{\SetFigFont{5}{6.0}{\rmdefault}{\mddefault}{\updefault}{\color[rgb]{0,0,0}$\vdots$}%
}}}}
\put(3916,-1636){\makebox(0,0)[lb]{\smash{{\SetFigFont{5}{6.0}{\rmdefault}{\mddefault}{\updefault}{\color[rgb]{0,0,0}$\vdots$}%
}}}}
\put(3916,-2221){\makebox(0,0)[lb]{\smash{{\SetFigFont{5}{6.0}{\rmdefault}{\mddefault}{\updefault}{\color[rgb]{0,0,0}$\vdots$}%
}}}}
\put(3916,-2806){\makebox(0,0)[lb]{\smash{{\SetFigFont{5}{6.0}{\rmdefault}{\mddefault}{\updefault}{\color[rgb]{0,0,0}$\vdots$}%
}}}}
\put(3916,-3976){\makebox(0,0)[lb]{\smash{{\SetFigFont{5}{6.0}{\rmdefault}{\mddefault}{\updefault}{\color[rgb]{0,0,0}$\vdots$}%
}}}}
\put(3916,-4561){\makebox(0,0)[lb]{\smash{{\SetFigFont{5}{6.0}{\rmdefault}{\mddefault}{\updefault}{\color[rgb]{0,0,0}$\vdots$}%
}}}}
\end{picture}%

%% file: rjdsc-fig1.pstex_t
\begin{picture}(0,0)%
\includegraphics{rjdsc-fig1.pstex}%
\end{picture}%
\setlength{\unitlength}{3315sp}%
\begingroup\makeatletter\ifx\SetFigFont\undefined%
\gdef\SetFigFont#1#2#3#4#5{%
  \reset@font\fontsize{#1}{#2pt}%
  \fontfamily{#3}\fontseries{#4}\fontshape{#5}%
  \selectfont}%
\fi\endgroup%
\begin{picture}(6504,3963)(-731,-3868)
\put(-314,-556){\makebox(0,0)[lb]{\smash{{\SetFigFont{6}{7.2}{\rmdefault}{\mddefault}{\updefault}{\color[rgb]{0,0,0}$z^{-1}$}%
}}}}
\put(-314,-1096){\makebox(0,0)[lb]{\smash{{\SetFigFont{6}{7.2}{\rmdefault}{\mddefault}{\updefault}{\color[rgb]{0,0,0}$z^{-1}$}%
}}}}
\put(-314,-2986){\makebox(0,0)[lb]{\smash{{\SetFigFont{6}{7.2}{\rmdefault}{\mddefault}{\updefault}{\color[rgb]{0,0,0}$z^{-1}$}%
}}}}
\put(1306,-736){\makebox(0,0)[lb]{\smash{{\SetFigFont{6}{7.2}{\rmdefault}{\mddefault}{\updefault}{\color[rgb]{0,0,0}$\getD{X_{1,k}}$}%
}}}}
\put(1306,-601){\makebox(0,0)[lb]{\smash{{\SetFigFont{6}{7.2}{\rmdefault}{\mddefault}{\updefault}{\color[rgb]{0,0,0}$\getC{X_{1,k}}$}%
}}}}
\put(1306,-466){\makebox(0,0)[lb]{\smash{{\SetFigFont{6}{7.2}{\rmdefault}{\mddefault}{\updefault}{\color[rgb]{0,0,0}$\getB{X_{1,k}}$}%
}}}}
\put(1306,-331){\makebox(0,0)[lb]{\smash{{\SetFigFont{6}{7.2}{\rmdefault}{\mddefault}{\updefault}{\color[rgb]{0,0,0}$\getA{X_{1,k}}$}%
}}}}
\put(1306,-151){\makebox(0,0)[lb]{\smash{{\SetFigFont{6}{7.2}{\rmdefault}{\mddefault}{\updefault}{\color[rgb]{0,0,0}$\getA{X_{0,k}}$}%
}}}}
\put(1306,-916){\makebox(0,0)[lb]{\smash{{\SetFigFont{6}{7.2}{\rmdefault}{\mddefault}{\updefault}{\color[rgb]{0,0,0}$\getA{X_{2,k}}$}%
}}}}
\put(1306,-1051){\makebox(0,0)[lb]{\smash{{\SetFigFont{6}{7.2}{\rmdefault}{\mddefault}{\updefault}{\color[rgb]{0,0,0}$\getD{X_{2,k}}$}%
}}}}
\put(1306,-1366){\makebox(0,0)[lb]{\smash{{\SetFigFont{6}{7.2}{\rmdefault}{\mddefault}{\updefault}{\color[rgb]{0,0,0}$\getB{X_{3,k}}$}%
}}}}
\put(1306,-1501){\makebox(0,0)[lb]{\smash{{\SetFigFont{6}{7.2}{\rmdefault}{\mddefault}{\updefault}{\color[rgb]{0,0,0}$\getC{X_{3,k}}$}%
}}}}
\put(1306,-1816){\makebox(0,0)[lb]{\smash{{\SetFigFont{6}{7.2}{\rmdefault}{\mddefault}{\updefault}{\color[rgb]{0,0,0}$\getA{X_{4,k}}$}%
}}}}
\put(1306,-1996){\makebox(0,0)[lb]{\smash{{\SetFigFont{6}{7.2}{\rmdefault}{\mddefault}{\updefault}{\color[rgb]{0,0,0}$\getA{X_{5,k}}$}%
}}}}
\put(1306,-2131){\makebox(0,0)[lb]{\smash{{\SetFigFont{6}{7.2}{\rmdefault}{\mddefault}{\updefault}{\color[rgb]{0,0,0}$\getB{X_{5,k}}$}%
}}}}
\put(1306,-2266){\makebox(0,0)[lb]{\smash{{\SetFigFont{6}{7.2}{\rmdefault}{\mddefault}{\updefault}{\color[rgb]{0,0,0}$\getC{X_{5,k}}$}%
}}}}
\put(1306,-2401){\makebox(0,0)[lb]{\smash{{\SetFigFont{6}{7.2}{\rmdefault}{\mddefault}{\updefault}{\color[rgb]{0,0,0}$\getD{X_{5,k}}$}%
}}}}
\put(1306,-2581){\makebox(0,0)[lb]{\smash{{\SetFigFont{6}{7.2}{\rmdefault}{\mddefault}{\updefault}{\color[rgb]{0,0,0}$\getA{X_{6,k}}$}%
}}}}
\put(1306,-2716){\makebox(0,0)[lb]{\smash{{\SetFigFont{6}{7.2}{\rmdefault}{\mddefault}{\updefault}{\color[rgb]{0,0,0}$\getD{X_{6,k}}$}%
}}}}
\put(1306,-1231){\makebox(0,0)[lb]{\smash{{\SetFigFont{6}{7.2}{\rmdefault}{\mddefault}{\updefault}{\color[rgb]{0,0,0}$\getA{X_{3,k}}$}%
}}}}
\put(1306,-2896){\makebox(0,0)[lb]{\smash{{\SetFigFont{6}{7.2}{\rmdefault}{\mddefault}{\updefault}{\color[rgb]{0,0,0}$\getA{X_{7,k}}$}%
}}}}
\put(1306,-3031){\makebox(0,0)[lb]{\smash{{\SetFigFont{6}{7.2}{\rmdefault}{\mddefault}{\updefault}{\color[rgb]{0,0,0}$\getB{X_{7,k}}$}%
}}}}
\put(1306,-3166){\makebox(0,0)[lb]{\smash{{\SetFigFont{6}{7.2}{\rmdefault}{\mddefault}{\updefault}{\color[rgb]{0,0,0}$\getC{X_{7,k}}$}%
}}}}
\put(1306,-3301){\makebox(0,0)[lb]{\smash{{\SetFigFont{6}{7.2}{\rmdefault}{\mddefault}{\updefault}{\color[rgb]{0,0,0}$\getD{X_{7,k}}$}%
}}}}
\put(721,-286){\makebox(0,0)[lb]{\smash{{\SetFigFont{6}{7.2}{\rmdefault}{\mddefault}{\updefault}{\color[rgb]{0,0,0}$x_{0,k}$}%
}}}}
\put(721,-3256){\makebox(0,0)[lb]{\smash{{\SetFigFont{6}{7.2}{\rmdefault}{\mddefault}{\updefault}{\color[rgb]{0,0,0}$x_{7,k}$}%
}}}}
\put(721,-826){\makebox(0,0)[lb]{\smash{{\SetFigFont{6}{7.2}{\rmdefault}{\mddefault}{\updefault}{\color[rgb]{0,0,0}$x_{1,k}$}%
}}}}
\put(721,-1366){\makebox(0,0)[lb]{\smash{{\SetFigFont{6}{7.2}{\rmdefault}{\mddefault}{\updefault}{\color[rgb]{0,0,0}$x_{2,k}$}%
}}}}
\put(721,-2716){\makebox(0,0)[lb]{\smash{{\SetFigFont{6}{7.2}{\rmdefault}{\mddefault}{\updefault}{\color[rgb]{0,0,0}$x_{6,k}$}%
}}}}
\put(2386,-106){\makebox(0,0)[lb]{\smash{{\SetFigFont{6}{7.2}{\rmdefault}{\mddefault}{\updefault}{\color[rgb]{0,0,0}$\getA{X_{0,k}}$}%
}}}}
\put(2386,-241){\makebox(0,0)[lb]{\smash{{\SetFigFont{6}{7.2}{\rmdefault}{\mddefault}{\updefault}{\color[rgb]{0,0,0}$\getA{X_{0,k-1}}$}%
}}}}
\put(2386,-736){\makebox(0,0)[lb]{\smash{{\SetFigFont{6}{7.2}{\rmdefault}{\mddefault}{\updefault}{\color[rgb]{0,0,0}$\getA{X_{0,k-7}}$}%
}}}}
\put(2386,-916){\makebox(0,0)[lb]{\smash{{\SetFigFont{6}{7.2}{\rmdefault}{\mddefault}{\updefault}{\color[rgb]{0,0,0}$\getA{X_{1,k}}$}%
}}}}
\put(2386,-1051){\makebox(0,0)[lb]{\smash{{\SetFigFont{6}{7.2}{\rmdefault}{\mddefault}{\updefault}{\color[rgb]{0,0,0}$\getA{X_{1,k-1}}$}%
}}}}
\put(2386,-1546){\makebox(0,0)[lb]{\smash{{\SetFigFont{6}{7.2}{\rmdefault}{\mddefault}{\updefault}{\color[rgb]{0,0,0}$\getA{X_{1,k-7}}$}%
}}}}
\put(2386,-1726){\makebox(0,0)[lb]{\smash{{\SetFigFont{6}{7.2}{\rmdefault}{\mddefault}{\updefault}{\color[rgb]{0,0,0}$\getB{X_{1,k}}$}%
}}}}
\put(2386,-2356){\makebox(0,0)[lb]{\smash{{\SetFigFont{6}{7.2}{\rmdefault}{\mddefault}{\updefault}{\color[rgb]{0,0,0}$\getB{X_{1,k-7}}$}%
}}}}
\put(2386,-2536){\makebox(0,0)[lb]{\smash{{\SetFigFont{6}{7.2}{\rmdefault}{\mddefault}{\updefault}{\color[rgb]{0,0,0}$\getD{X_{7,k}}$}%
}}}}
\put(2386,-2671){\makebox(0,0)[lb]{\smash{{\SetFigFont{6}{7.2}{\rmdefault}{\mddefault}{\updefault}{\color[rgb]{0,0,0}$\getD{X_{7,k-1}}$}%
}}}}
\put(2386,-3211){\makebox(0,0)[lb]{\smash{{\SetFigFont{6}{7.2}{\rmdefault}{\mddefault}{\updefault}{\color[rgb]{0,0,0}$\getD{X_{7,k-7}}$}%
}}}}
\put(2386,-1861){\makebox(0,0)[lb]{\smash{{\SetFigFont{6}{7.2}{\rmdefault}{\mddefault}{\updefault}{\color[rgb]{0,0,0}$\getB{X_{1,k-1}}$}%
}}}}
\put(2701,-2986){\makebox(0,0)[b]{\smash{{\SetFigFont{10}{12.0}{\rmdefault}{\mddefault}{\updefault}{\color[rgb]{0,0,0}$\vdots$}%
}}}}
\put(2701,-2176){\makebox(0,0)[b]{\smash{{\SetFigFont{10}{12.0}{\rmdefault}{\mddefault}{\updefault}{\color[rgb]{0,0,0}$\vdots$}%
}}}}
\put(2701,-1366){\makebox(0,0)[b]{\smash{{\SetFigFont{10}{12.0}{\rmdefault}{\mddefault}{\updefault}{\color[rgb]{0,0,0}$\vdots$}%
}}}}
\put(2701,-556){\makebox(0,0)[b]{\smash{{\SetFigFont{10}{12.0}{\rmdefault}{\mddefault}{\updefault}{\color[rgb]{0,0,0}$\vdots$}%
}}}}
\put(5491,-241){\makebox(0,0)[lb]{\smash{{\SetFigFont{6}{7.2}{\rmdefault}{\mddefault}{\updefault}{\color[rgb]{0,0,0}$X_{0,k}^{(2D)}$}%
}}}}
\put(5491,-646){\makebox(0,0)[lb]{\smash{{\SetFigFont{6}{7.2}{\rmdefault}{\mddefault}{\updefault}{\color[rgb]{0,0,0}$X_{1,k}^{(2D)}$}%
}}}}
\put(5491,-1051){\makebox(0,0)[lb]{\smash{{\SetFigFont{6}{7.2}{\rmdefault}{\mddefault}{\updefault}{\color[rgb]{0,0,0}$X_{2,k}^{(2D)}$}%
}}}}
\put(5491,-1456){\makebox(0,0)[lb]{\smash{{\SetFigFont{6}{7.2}{\rmdefault}{\mddefault}{\updefault}{\color[rgb]{0,0,0}$X_{3,k}^{(2D)}$}%
}}}}
\put(5491,-1861){\makebox(0,0)[lb]{\smash{{\SetFigFont{6}{7.2}{\rmdefault}{\mddefault}{\updefault}{\color[rgb]{0,0,0}$X_{4,k}^{(2D)}$}%
}}}}
\put(5491,-2266){\makebox(0,0)[lb]{\smash{{\SetFigFont{6}{7.2}{\rmdefault}{\mddefault}{\updefault}{\color[rgb]{0,0,0}$X_{5,k}^{(2D)}$}%
}}}}
\put(5491,-3076){\makebox(0,0)[lb]{\smash{{\SetFigFont{6}{7.2}{\rmdefault}{\mddefault}{\updefault}{\color[rgb]{0,0,0}$X_{7,k}^{(2D)}$}%
}}}}
\put(5491,-2671){\makebox(0,0)[lb]{\smash{{\SetFigFont{6}{7.2}{\rmdefault}{\mddefault}{\updefault}{\color[rgb]{0,0,0}$X_{6,k}^{(2D)}$}%
}}}}
\put(1306,-1636){\makebox(0,0)[lb]{\smash{{\SetFigFont{6}{7.2}{\rmdefault}{\mddefault}{\updefault}{\color[rgb]{0,0,0}$\getD{X_{3,k}}$}%
}}}}
\put(-179,-16){\makebox(0,0)[b]{\smash{{\SetFigFont{6}{7.2}{\rmdefault}{\mddefault}{\updefault}{\color[rgb]{0,0,0}Input Sequence @ $F_\text{s}$}%
}}}}
\put(496,-196){\makebox(0,0)[b]{\smash{{\SetFigFont{6}{7.2}{\rmdefault}{\mddefault}{\updefault}{\color[rgb]{0,0,0}@$F_\text{clock}$}%
}}}}
\put(-179,-2041){\makebox(0,0)[b]{\smash{{\SetFigFont{10}{12.0}{\rmdefault}{\mddefault}{\updefault}{\color[rgb]{0,0,0}$\vdots$}%
}}}}
\put(226,-2041){\makebox(0,0)[b]{\smash{{\SetFigFont{10}{12.0}{\rmdefault}{\mddefault}{\updefault}{\color[rgb]{0,0,0}$\vdots$}%
}}}}
\end{picture}%

%% file: rjdsc-fig2.pstex_t
\begin{picture}(0,0)%
\includegraphics{rjdsc-fig2.pstex}%
\end{picture}%
\setlength{\unitlength}{3315sp}%
\begingroup\makeatletter\ifx\SetFigFont\undefined%
\gdef\SetFigFont#1#2#3#4#5{%
  \reset@font\fontsize{#1}{#2pt}%
  \fontfamily{#3}\fontseries{#4}\fontshape{#5}%
  \selectfont}%
\fi\endgroup%
\begin{picture}(2232,4392)(2371,-3991)
\put(2386,-916){\makebox(0,0)[lb]{\smash{{\SetFigFont{6}{7.2}{\rmdefault}{\mddefault}{\updefault}{\color[rgb]{0,0,0}$\getQ{X_{1,0}}$}%
}}}}
\put(2386,-1051){\makebox(0,0)[lb]{\smash{{\SetFigFont{6}{7.2}{\rmdefault}{\mddefault}{\updefault}{\color[rgb]{0,0,0}$\getQ{X_{1,1}}$}%
}}}}
\put(2386,-1546){\makebox(0,0)[lb]{\smash{{\SetFigFont{6}{7.2}{\rmdefault}{\mddefault}{\updefault}{\color[rgb]{0,0,0}$\getQ{X_{1,7}}$}%
}}}}
\put(2701,-1366){\makebox(0,0)[b]{\smash{{\SetFigFont{10}{12.0}{\rmdefault}{\mddefault}{\updefault}{\color[rgb]{0,0,0}$\vdots$}%
}}}}
\put(2386,119){\makebox(0,0)[lb]{\smash{{\SetFigFont{6}{7.2}{\rmdefault}{\mddefault}{\updefault}{\color[rgb]{0,0,0}$\getQ{X_{0,0}}$}%
}}}}
\put(2386,-16){\makebox(0,0)[lb]{\smash{{\SetFigFont{6}{7.2}{\rmdefault}{\mddefault}{\updefault}{\color[rgb]{0,0,0}$\getQ{X_{0,1}}$}%
}}}}
\put(2386,-511){\makebox(0,0)[lb]{\smash{{\SetFigFont{6}{7.2}{\rmdefault}{\mddefault}{\updefault}{\color[rgb]{0,0,0}$\getQ{X_{0,7}}$}%
}}}}
\put(2701,-331){\makebox(0,0)[b]{\smash{{\SetFigFont{10}{12.0}{\rmdefault}{\mddefault}{\updefault}{\color[rgb]{0,0,0}$\vdots$}%
}}}}
\put(2386,-3031){\makebox(0,0)[lb]{\smash{{\SetFigFont{6}{7.2}{\rmdefault}{\mddefault}{\updefault}{\color[rgb]{0,0,0}$\getQ{X_{7,0}}$}%
}}}}
\put(2386,-3166){\makebox(0,0)[lb]{\smash{{\SetFigFont{6}{7.2}{\rmdefault}{\mddefault}{\updefault}{\color[rgb]{0,0,0}$\getQ{X_{7,1}}$}%
}}}}
\put(2386,-3661){\makebox(0,0)[lb]{\smash{{\SetFigFont{6}{7.2}{\rmdefault}{\mddefault}{\updefault}{\color[rgb]{0,0,0}$\getQ{X_{7,7}}$}%
}}}}
\put(2701,-3481){\makebox(0,0)[b]{\smash{{\SetFigFont{10}{12.0}{\rmdefault}{\mddefault}{\updefault}{\color[rgb]{0,0,0}$\vdots$}%
}}}}
\put(3736,-196){\makebox(0,0)[lb]{\smash{{\SetFigFont{6}{7.2}{\rmdefault}{\mddefault}{\updefault}{\color[rgb]{0,0,0}$\getQ{x_{0,k}}$}%
}}}}
\put(3736,-1231){\makebox(0,0)[lb]{\smash{{\SetFigFont{6}{7.2}{\rmdefault}{\mddefault}{\updefault}{\color[rgb]{0,0,0}$\getQ{x_{1,k}}$}%
}}}}
\put(3736,-3346){\makebox(0,0)[lb]{\smash{{\SetFigFont{6}{7.2}{\rmdefault}{\mddefault}{\updefault}{\color[rgb]{0,0,0}$\getQ{x_{7,k}}$}%
}}}}
\put(4546,254){\makebox(0,0)[lb]{\smash{{\SetFigFont{6}{7.2}{\rmdefault}{\mddefault}{\updefault}{\color[rgb]{0,0,0}$\getQ{\getA{X_{0,k}}}$}%
}}}}
\put(4546, 74){\makebox(0,0)[lb]{\smash{{\SetFigFont{6}{7.2}{\rmdefault}{\mddefault}{\updefault}{\color[rgb]{0,0,0}$\getQ{\getA{X_{1,k}}}$}%
}}}}
\put(4546,-106){\makebox(0,0)[lb]{\smash{{\SetFigFont{6}{7.2}{\rmdefault}{\mddefault}{\updefault}{\color[rgb]{0,0,0}$\getQ{\getB{X_{1,k}}}$}%
}}}}
\put(4546,-286){\makebox(0,0)[lb]{\smash{{\SetFigFont{6}{7.2}{\rmdefault}{\mddefault}{\updefault}{\color[rgb]{0,0,0}$\getQ{\getC{X_{1,k}}}$}%
}}}}
\put(4546,-466){\makebox(0,0)[lb]{\smash{{\SetFigFont{6}{7.2}{\rmdefault}{\mddefault}{\updefault}{\color[rgb]{0,0,0}$\getQ{\getD{X_{1,k}}}$}%
}}}}
\put(4546,-646){\makebox(0,0)[lb]{\smash{{\SetFigFont{6}{7.2}{\rmdefault}{\mddefault}{\updefault}{\color[rgb]{0,0,0}$\getQ{\getA{X_{2,k}}}$}%
}}}}
\put(4546,-826){\makebox(0,0)[lb]{\smash{{\SetFigFont{6}{7.2}{\rmdefault}{\mddefault}{\updefault}{\color[rgb]{0,0,0}$\getQ{\getD{X_{2,k}}}$}%
}}}}
\put(4546,-1006){\makebox(0,0)[lb]{\smash{{\SetFigFont{6}{7.2}{\rmdefault}{\mddefault}{\updefault}{\color[rgb]{0,0,0}$\getQ{\getA{X_{3,k}}}$}%
}}}}
\put(4546,-1186){\makebox(0,0)[lb]{\smash{{\SetFigFont{6}{7.2}{\rmdefault}{\mddefault}{\updefault}{\color[rgb]{0,0,0}$\getQ{\getB{X_{3,k}}}$}%
}}}}
\put(4546,-1366){\makebox(0,0)[lb]{\smash{{\SetFigFont{6}{7.2}{\rmdefault}{\mddefault}{\updefault}{\color[rgb]{0,0,0}$\getQ{\getC{X_{3,k}}}$}%
}}}}
\put(4546,-1546){\makebox(0,0)[lb]{\smash{{\SetFigFont{6}{7.2}{\rmdefault}{\mddefault}{\updefault}{\color[rgb]{0,0,0}$\getQ{\getD{X_{3,k}}}$}%
}}}}
\put(4546,-1726){\makebox(0,0)[lb]{\smash{{\SetFigFont{6}{7.2}{\rmdefault}{\mddefault}{\updefault}{\color[rgb]{0,0,0}$\getQ{\getA{X_{4,k}}}$}%
}}}}
\put(4546,-1906){\makebox(0,0)[lb]{\smash{{\SetFigFont{6}{7.2}{\rmdefault}{\mddefault}{\updefault}{\color[rgb]{0,0,0}$\getQ{\getA{X_{5,k}}}$}%
}}}}
\put(4546,-2086){\makebox(0,0)[lb]{\smash{{\SetFigFont{6}{7.2}{\rmdefault}{\mddefault}{\updefault}{\color[rgb]{0,0,0}$\getQ{\getB{X_{5,k}}}$}%
}}}}
\put(4546,-2266){\makebox(0,0)[lb]{\smash{{\SetFigFont{6}{7.2}{\rmdefault}{\mddefault}{\updefault}{\color[rgb]{0,0,0}$\getQ{\getC{X_{5,k}}}$}%
}}}}
\put(4546,-2446){\makebox(0,0)[lb]{\smash{{\SetFigFont{6}{7.2}{\rmdefault}{\mddefault}{\updefault}{\color[rgb]{0,0,0}$\getQ{\getD{X_{5,k}}}$}%
}}}}
\put(4546,-2626){\makebox(0,0)[lb]{\smash{{\SetFigFont{6}{7.2}{\rmdefault}{\mddefault}{\updefault}{\color[rgb]{0,0,0}$\getQ{\getA{X_{6,k}}}$}%
}}}}
\put(4546,-2806){\makebox(0,0)[lb]{\smash{{\SetFigFont{6}{7.2}{\rmdefault}{\mddefault}{\updefault}{\color[rgb]{0,0,0}$\getQ{\getD{X_{6,k}}}$}%
}}}}
\put(4546,-2986){\makebox(0,0)[lb]{\smash{{\SetFigFont{6}{7.2}{\rmdefault}{\mddefault}{\updefault}{\color[rgb]{0,0,0}$\getQ{\getA{X_{7,k}}}$}%
}}}}
\put(4546,-3166){\makebox(0,0)[lb]{\smash{{\SetFigFont{6}{7.2}{\rmdefault}{\mddefault}{\updefault}{\color[rgb]{0,0,0}$\getQ{\getB{X_{7,k}}}$}%
}}}}
\put(4546,-3346){\makebox(0,0)[lb]{\smash{{\SetFigFont{6}{7.2}{\rmdefault}{\mddefault}{\updefault}{\color[rgb]{0,0,0}$\getQ{\getC{X_{7,k}}}$}%
}}}}
\put(4546,-3526){\makebox(0,0)[lb]{\smash{{\SetFigFont{6}{7.2}{\rmdefault}{\mddefault}{\updefault}{\color[rgb]{0,0,0}$\getQ{\getD{X_{7,k}}}$}%
}}}}
\put(3241,-2221){\makebox(0,0)[b]{\smash{{\SetFigFont{10}{12.0}{\rmdefault}{\mddefault}{\updefault}{\color[rgb]{0,0,0}$\vdots$}%
}}}}
\end{picture}%

%% file: rjdsc-fig-6.pstex_t
\begin{picture}(0,0)%
\includegraphics{rjdsc-fig-6.pstex}%
\end{picture}%
\setlength{\unitlength}{2901sp}%
\begingroup\makeatletter\ifx\SetFigFont\undefined%
\gdef\SetFigFont#1#2#3#4#5{%
  \reset@font\fontsize{#1}{#2pt}%
  \fontfamily{#3}\fontseries{#4}\fontshape{#5}%
  \selectfont}%
\fi\endgroup%
\begin{picture}(2817,3624)(2056,-3448)
\put(2071, 29){\makebox(0,0)[rb]{\smash{{\SetFigFont{7}{8.4}{\rmdefault}{\mddefault}{\updefault}{\color[rgb]{0,0,0}$\getA{\getA{X_{i,k}}}$}%
}}}}
\put(2071,-196){\makebox(0,0)[rb]{\smash{{\SetFigFont{7}{8.4}{\rmdefault}{\mddefault}{\updefault}{\color[rgb]{0,0,0}$\getB{\getA{X_{i,k}}}$}%
}}}}
\put(2071,-421){\makebox(0,0)[rb]{\smash{{\SetFigFont{7}{8.4}{\rmdefault}{\mddefault}{\updefault}{\color[rgb]{0,0,0}$\getC{\getA{X_{i,k}}}$}%
}}}}
\put(2071,-646){\makebox(0,0)[rb]{\smash{{\SetFigFont{7}{8.4}{\rmdefault}{\mddefault}{\updefault}{\color[rgb]{0,0,0}$\getD{\getA{X_{i,k}}}$}%
}}}}
\put(2071,-871){\makebox(0,0)[rb]{\smash{{\SetFigFont{7}{8.4}{\rmdefault}{\mddefault}{\updefault}{\color[rgb]{0,0,0}$\getA{\getB{X_{i,k}}}$}%
}}}}
\put(2071,-1096){\makebox(0,0)[rb]{\smash{{\SetFigFont{7}{8.4}{\rmdefault}{\mddefault}{\updefault}{\color[rgb]{0,0,0}$\getB{\getB{X_{i,k}}}$}%
}}}}
\put(2071,-1321){\makebox(0,0)[rb]{\smash{{\SetFigFont{7}{8.4}{\rmdefault}{\mddefault}{\updefault}{\color[rgb]{0,0,0}$\getC{\getB{X_{i,k}}}$}%
}}}}
\put(2071,-1546){\makebox(0,0)[rb]{\smash{{\SetFigFont{7}{8.4}{\rmdefault}{\mddefault}{\updefault}{\color[rgb]{0,0,0}$\getD{\getB{X_{i,k}}}$}%
}}}}
\put(2071,-1771){\makebox(0,0)[rb]{\smash{{\SetFigFont{7}{8.4}{\rmdefault}{\mddefault}{\updefault}{\color[rgb]{0,0,0}$\getA{\getC{X_{i,k}}}$}%
}}}}
\put(2071,-1996){\makebox(0,0)[rb]{\smash{{\SetFigFont{7}{8.4}{\rmdefault}{\mddefault}{\updefault}{\color[rgb]{0,0,0}$\getB{\getC{X_{i,k}}}$}%
}}}}
\put(2071,-2221){\makebox(0,0)[rb]{\smash{{\SetFigFont{7}{8.4}{\rmdefault}{\mddefault}{\updefault}{\color[rgb]{0,0,0}$\getC{\getC{X_{i,k}}}$}%
}}}}
\put(2071,-2446){\makebox(0,0)[rb]{\smash{{\SetFigFont{7}{8.4}{\rmdefault}{\mddefault}{\updefault}{\color[rgb]{0,0,0}$\getD{\getC{X_{i,k}}}$}%
}}}}
\put(2071,-2671){\makebox(0,0)[rb]{\smash{{\SetFigFont{7}{8.4}{\rmdefault}{\mddefault}{\updefault}{\color[rgb]{0,0,0}$\getA{\getD{X_{i,k}}}$}%
}}}}
\put(2071,-2896){\makebox(0,0)[rb]{\smash{{\SetFigFont{7}{8.4}{\rmdefault}{\mddefault}{\updefault}{\color[rgb]{0,0,0}$\getB{\getD{X_{i,k}}}$}%
}}}}
\put(2071,-3121){\makebox(0,0)[rb]{\smash{{\SetFigFont{7}{8.4}{\rmdefault}{\mddefault}{\updefault}{\color[rgb]{0,0,0}$\getC{\getD{X_{i,k}}}$}%
}}}}
\put(2071,-3346){\makebox(0,0)[rb]{\smash{{\SetFigFont{7}{8.4}{\rmdefault}{\mddefault}{\updefault}{\color[rgb]{0,0,0}$\getD{\getD{X_{i,k}}}$}%
}}}}
\put(4771,-1546){\makebox(0,0)[lb]{\smash{{\SetFigFont{7}{8.4}{\rmdefault}{\mddefault}{\updefault}{\color[rgb]{0,0,0}$\alpha \cdot X_{i,k}$}%
}}}}
\put(3736,-151){\makebox(0,0)[b]{\smash{{\SetFigFont{7}{8.4}{\rmdefault}{\mddefault}{\updefault}{\color[rgb]{0,0,0}Booth }%
}}}}
\put(3736,-376){\makebox(0,0)[b]{\smash{{\SetFigFont{7}{8.4}{\rmdefault}{\mddefault}{\updefault}{\color[rgb]{0,0,0}encoded}%
}}}}
\put(3736,-601){\makebox(0,0)[b]{\smash{{\SetFigFont{7}{8.4}{\rmdefault}{\mddefault}{\updefault}{\color[rgb]{0,0,0}$\alpha$}%
}}}}
\put(3421,-61){\makebox(0,0)[rb]{\smash{{\SetFigFont{7}{8.4}{\rmdefault}{\mddefault}{\updefault}{\color[rgb]{0,0,0}$\getA{Y_{i,k}}$}%
}}}}
\put(3421,-1051){\makebox(0,0)[rb]{\smash{{\SetFigFont{7}{8.4}{\rmdefault}{\mddefault}{\updefault}{\color[rgb]{0,0,0}$\getB{Y_{i,k}}$}%
}}}}
\put(3421,-2041){\makebox(0,0)[rb]{\smash{{\SetFigFont{7}{8.4}{\rmdefault}{\mddefault}{\updefault}{\color[rgb]{0,0,0}$\getC{Y_{i,k}}$}%
}}}}
\put(3421,-3031){\makebox(0,0)[rb]{\smash{{\SetFigFont{7}{8.4}{\rmdefault}{\mddefault}{\updefault}{\color[rgb]{0,0,0}$\getD{Y_{i,k}}$}%
}}}}
\end{picture}%

%% file: fpga_results_4.pstex_t
\begin{picture}(0,0)%
\includegraphics{fpga_results_4.pstex}%
\end{picture}%
\setlength{\unitlength}{3355sp}%
\begingroup\makeatletter\ifx\SetFigFont\undefined%
\gdef\SetFigFont#1#2#3#4#5{%
  \reset@font\fontsize{#1}{#2pt}%
  \fontfamily{#3}\fontseries{#4}\fontshape{#5}%
  \selectfont}%
\fi\endgroup%
\begin{picture}(3323,6072)(586,-5728)
\put(1801,-451){\makebox(0,0)[lb]{\smash{{\SetFigFont{7}{8.4}{\rmdefault}{\mddefault}{\updefault}Slice registers}}}}
\put(601,-676){\makebox(0,0)[lb]{\smash{{\SetFigFont{7}{8.4}{\rmdefault}{\mddefault}{\updefault}Design 1}}}}
\put(601,-826){\makebox(0,0)[lb]{\smash{{\SetFigFont{7}{8.4}{\rmdefault}{\mddefault}{\updefault}Design 3}}}}
\put(601,-976){\makebox(0,0)[lb]{\smash{{\SetFigFont{7}{8.4}{\rmdefault}{\mddefault}{\updefault}Design 5}}}}
\put(1801,-1096){\makebox(0,0)[lb]{\smash{{\SetFigFont{7}{8.4}{\rmdefault}{\mddefault}{\updefault}Slice LUTs}}}}
\put(601,-1321){\makebox(0,0)[lb]{\smash{{\SetFigFont{7}{8.4}{\rmdefault}{\mddefault}{\updefault}Design 1}}}}
\put(601,-1471){\makebox(0,0)[lb]{\smash{{\SetFigFont{7}{8.4}{\rmdefault}{\mddefault}{\updefault}Design 3}}}}
\put(601,-1621){\makebox(0,0)[lb]{\smash{{\SetFigFont{7}{8.4}{\rmdefault}{\mddefault}{\updefault}Design 5}}}}
\put(3479,-784){\makebox(0,0)[lb]{\smash{{\SetFigFont{7}{8.4}{\rmdefault}{\mddefault}{\updefault}7104}}}}
\put(3496,-1284){\makebox(0,0)[lb]{\smash{{\SetFigFont{7}{8.4}{\rmdefault}{\mddefault}{\updefault}9628}}}}
\put(3021,-643){\makebox(0,0)[lb]{\smash{{\SetFigFont{7}{8.4}{\rmdefault}{\mddefault}{\updefault}5767}}}}
\put(3421,-946){\makebox(0,0)[lb]{\smash{{\SetFigFont{7}{8.4}{\rmdefault}{\mddefault}{\updefault}7286}}}}
\put(3083,-1442){\makebox(0,0)[lb]{\smash{{\SetFigFont{7}{8.4}{\rmdefault}{\mddefault}{\updefault}7784}}}}
\put(3305,-1591){\makebox(0,0)[lb]{\smash{{\SetFigFont{7}{8.4}{\rmdefault}{\mddefault}{\updefault}8839}}}}
\put(1801,239){\makebox(0,0)[lb]{\smash{{\SetFigFont{7}{8.4}{\rmdefault}{\mddefault}{\updefault}Slices}}}}
\put(601, 14){\makebox(0,0)[lb]{\smash{{\SetFigFont{7}{8.4}{\rmdefault}{\mddefault}{\updefault}Design 1}}}}
\put(601,-136){\makebox(0,0)[lb]{\smash{{\SetFigFont{7}{8.4}{\rmdefault}{\mddefault}{\updefault}Design 3}}}}
\put(601,-286){\makebox(0,0)[lb]{\smash{{\SetFigFont{7}{8.4}{\rmdefault}{\mddefault}{\updefault}Design 5}}}}
\put(3478, 42){\makebox(0,0)[lb]{\smash{{\SetFigFont{7}{8.4}{\rmdefault}{\mddefault}{\updefault}2818}}}}
\put(3119,-111){\makebox(0,0)[lb]{\smash{{\SetFigFont{7}{8.4}{\rmdefault}{\mddefault}{\updefault}2377}}}}
\put(3314,-269){\makebox(0,0)[lb]{\smash{{\SetFigFont{7}{8.4}{\rmdefault}{\mddefault}{\updefault}2605}}}}
\put(1801,-1801){\makebox(0,0)[lb]{\smash{{\SetFigFont{7}{8.4}{\rmdefault}{\mddefault}{\updefault}Frequency (MHz)}}}}
\put(601,-4741){\makebox(0,0)[lb]{\smash{{\SetFigFont{7}{8.4}{\rmdefault}{\mddefault}{\updefault}Design 1}}}}
\put(601,-4891){\makebox(0,0)[lb]{\smash{{\SetFigFont{7}{8.4}{\rmdefault}{\mddefault}{\updefault}Design 3}}}}
\put(601,-5041){\makebox(0,0)[lb]{\smash{{\SetFigFont{7}{8.4}{\rmdefault}{\mddefault}{\updefault}Design 5}}}}
\put(601,-2026){\makebox(0,0)[lb]{\smash{{\SetFigFont{7}{8.4}{\rmdefault}{\mddefault}{\updefault}Design 1}}}}
\put(601,-2176){\makebox(0,0)[lb]{\smash{{\SetFigFont{7}{8.4}{\rmdefault}{\mddefault}{\updefault}Design 3}}}}
\put(601,-2326){\makebox(0,0)[lb]{\smash{{\SetFigFont{7}{8.4}{\rmdefault}{\mddefault}{\updefault}Design 5}}}}
\put(1801,-2491){\makebox(0,0)[lb]{\smash{{\SetFigFont{7}{8.4}{\rmdefault}{\mddefault}{\updefault}Quies. power (W)}}}}
\put(601,-2716){\makebox(0,0)[lb]{\smash{{\SetFigFont{7}{8.4}{\rmdefault}{\mddefault}{\updefault}Design 1}}}}
\put(601,-2866){\makebox(0,0)[lb]{\smash{{\SetFigFont{7}{8.4}{\rmdefault}{\mddefault}{\updefault}Design 3}}}}
\put(601,-3016){\makebox(0,0)[lb]{\smash{{\SetFigFont{7}{8.4}{\rmdefault}{\mddefault}{\updefault}Design 5}}}}
\put(1951,-3166){\makebox(0,0)[lb]{\smash{{\SetFigFont{7}{8.4}{\rmdefault}{\mddefault}{\updefault}Dyn. power (W)}}}}
\put(601,-3391){\makebox(0,0)[lb]{\smash{{\SetFigFont{7}{8.4}{\rmdefault}{\mddefault}{\updefault}Design 1}}}}
\put(601,-3541){\makebox(0,0)[lb]{\smash{{\SetFigFont{7}{8.4}{\rmdefault}{\mddefault}{\updefault}Design 3}}}}
\put(601,-3691){\makebox(0,0)[lb]{\smash{{\SetFigFont{7}{8.4}{\rmdefault}{\mddefault}{\updefault}Design 5}}}}
\put(1951,-3841){\makebox(0,0)[lb]{\smash{{\SetFigFont{7}{8.4}{\rmdefault}{\mddefault}{\updefault}Total power (W)}}}}
\put(601,-4066){\makebox(0,0)[lb]{\smash{{\SetFigFont{7}{8.4}{\rmdefault}{\mddefault}{\updefault}Design 1}}}}
\put(601,-4216){\makebox(0,0)[lb]{\smash{{\SetFigFont{7}{8.4}{\rmdefault}{\mddefault}{\updefault}Design 3}}}}
\put(601,-4366){\makebox(0,0)[lb]{\smash{{\SetFigFont{7}{8.4}{\rmdefault}{\mddefault}{\updefault}Design 5}}}}
\put(601,-5341){\makebox(0,0)[lb]{\smash{{\SetFigFont{7}{8.4}{\rmdefault}{\mddefault}{\updefault}Design 1}}}}
\put(601,-5491){\makebox(0,0)[lb]{\smash{{\SetFigFont{7}{8.4}{\rmdefault}{\mddefault}{\updefault}Design 3}}}}
\put(601,-5641){\makebox(0,0)[lb]{\smash{{\SetFigFont{7}{8.4}{\rmdefault}{\mddefault}{\updefault}Design 5}}}}
\put(2189,-1978){\makebox(0,0)[lb]{\smash{{\SetFigFont{7}{8.4}{\rmdefault}{\mddefault}{\updefault}130.41}}}}
\put(3443,-2138){\makebox(0,0)[lb]{\smash{{\SetFigFont{7}{8.4}{\rmdefault}{\mddefault}{\updefault}309.885}}}}
\put(3495,-2300){\makebox(0,0)[lb]{\smash{{\SetFigFont{7}{8.4}{\rmdefault}{\mddefault}{\updefault}312.402}}}}
\put(3478,-2658){\makebox(0,0)[lb]{\smash{{\SetFigFont{7}{8.4}{\rmdefault}{\mddefault}{\updefault}2.740}}}}
\put(3493,-2832){\makebox(0,0)[lb]{\smash{{\SetFigFont{7}{8.4}{\rmdefault}{\mddefault}{\updefault}2.773}}}}
\put(3448,-2982){\makebox(0,0)[lb]{\smash{{\SetFigFont{7}{8.4}{\rmdefault}{\mddefault}{\updefault}2.740}}}}
\put(2330,-3344){\makebox(0,0)[lb]{\smash{{\SetFigFont{7}{8.4}{\rmdefault}{\mddefault}{\updefault}0.897}}}}
\put(3476,-3502){\makebox(0,0)[lb]{\smash{{\SetFigFont{7}{8.4}{\rmdefault}{\mddefault}{\updefault}1.871}}}}
\put(2365,-3668){\makebox(0,0)[lb]{\smash{{\SetFigFont{7}{8.4}{\rmdefault}{\mddefault}{\updefault}0.912}}}}
\put(2996,-4021){\makebox(0,0)[lb]{\smash{{\SetFigFont{7}{8.4}{\rmdefault}{\mddefault}{\updefault}3.637}}}}
\put(3491,-4181){\makebox(0,0)[lb]{\smash{{\SetFigFont{7}{8.4}{\rmdefault}{\mddefault}{\updefault}4.643}}}}
\put(3023,-4346){\makebox(0,0)[lb]{\smash{{\SetFigFont{7}{8.4}{\rmdefault}{\mddefault}{\updefault}3.652}}}}
\put(3495,-4706){\makebox(0,0)[lb]{\smash{{\SetFigFont{7}{8.4}{\rmdefault}{\mddefault}{\updefault}21.61}}}}
\put(2061,-4864){\makebox(0,0)[lb]{\smash{{\SetFigFont{7}{8.4}{\rmdefault}{\mddefault}{\updefault}7.67}}}}
\put(2105,-5003){\makebox(0,0)[lb]{\smash{{\SetFigFont{7}{8.4}{\rmdefault}{\mddefault}{\updefault}8.34}}}}
\put(3505,-5383){\makebox(0,0)[lb]{\smash{{\SetFigFont{7}{8.4}{\rmdefault}{\mddefault}{\updefault}0.213}}}}
\put(1499,-5530){\makebox(0,0)[lb]{\smash{{\SetFigFont{7}{8.4}{\rmdefault}{\mddefault}{\updefault}0.025}}}}
\put(1537,-5679){\makebox(0,0)[lb]{\smash{{\SetFigFont{7}{8.4}{\rmdefault}{\mddefault}{\updefault}0.028}}}}
\put(1951,-4516){\makebox(0,0)[lb]{\smash{{\SetFigFont{7}{8.4}{\rmdefault}{\mddefault}{\updefault}$\text{area}\times\text{time}$ ($\text{slices}\cdot\mu\mathrm{s}$)}}}}
\put(1951,-5191){\makebox(0,0)[lb]{\smash{{\SetFigFont{7}{8.4}{\rmdefault}{\mddefault}{\updefault}$\text{area}\times\text{time}^2$ ($\text{slices}\cdot\mu\mathrm{s}^2$)}}}}
\end{picture}%

%% file: fpga_results_8.pstex_t
\begin{picture}(0,0)%
\includegraphics{fpga_results_8.pstex}%
\end{picture}%
\setlength{\unitlength}{3355sp}%
\begingroup\makeatletter\ifx\SetFigFont\undefined%
\gdef\SetFigFont#1#2#3#4#5{%
  \reset@font\fontsize{#1}{#2pt}%
  \fontfamily{#3}\fontseries{#4}\fontshape{#5}%
  \selectfont}%
\fi\endgroup%
\begin{picture}(3314,6076)(586,-5882)
\put(601,-136){\makebox(0,0)[lb]{\smash{{\SetFigFont{7}{8.4}{\rmdefault}{\mddefault}{\updefault}Design 2}}}}
\put(601,-286){\makebox(0,0)[lb]{\smash{{\SetFigFont{7}{8.4}{\rmdefault}{\mddefault}{\updefault}Design 4}}}}
\put(601,-436){\makebox(0,0)[lb]{\smash{{\SetFigFont{7}{8.4}{\rmdefault}{\mddefault}{\updefault}Design 6}}}}
\put(1801, 89){\makebox(0,0)[lb]{\smash{{\SetFigFont{7}{8.4}{\rmdefault}{\mddefault}{\updefault}Slices}}}}
\put(3503,-97){\makebox(0,0)[lb]{\smash{{\SetFigFont{7}{8.4}{\rmdefault}{\mddefault}{\updefault}3618}}}}
\put(3379,-410){\makebox(0,0)[lb]{\smash{{\SetFigFont{7}{8.4}{\rmdefault}{\mddefault}{\updefault}3445}}}}
\put(601,-811){\makebox(0,0)[lb]{\smash{{\SetFigFont{7}{8.4}{\rmdefault}{\mddefault}{\updefault}Design 2}}}}
\put(601,-961){\makebox(0,0)[lb]{\smash{{\SetFigFont{7}{8.4}{\rmdefault}{\mddefault}{\updefault}Design 4}}}}
\put(601,-1111){\makebox(0,0)[lb]{\smash{{\SetFigFont{7}{8.4}{\rmdefault}{\mddefault}{\updefault}Design 6}}}}
\put(1801,-586){\makebox(0,0)[lb]{\smash{{\SetFigFont{7}{8.4}{\rmdefault}{\mddefault}{\updefault}Slice registers}}}}
\put(601,-1486){\makebox(0,0)[lb]{\smash{{\SetFigFont{7}{8.4}{\rmdefault}{\mddefault}{\updefault}Design 2}}}}
\put(601,-1636){\makebox(0,0)[lb]{\smash{{\SetFigFont{7}{8.4}{\rmdefault}{\mddefault}{\updefault}Design 4}}}}
\put(601,-1786){\makebox(0,0)[lb]{\smash{{\SetFigFont{7}{8.4}{\rmdefault}{\mddefault}{\updefault}Design 6}}}}
\put(1801,-1261){\makebox(0,0)[lb]{\smash{{\SetFigFont{7}{8.4}{\rmdefault}{\mddefault}{\updefault}Slice LUTs}}}}
\put(3472,-1448){\makebox(0,0)[lb]{\smash{{\SetFigFont{7}{8.4}{\rmdefault}{\mddefault}{\updefault}12794}}}}
\put(3046,-1607){\makebox(0,0)[lb]{\smash{{\SetFigFont{7}{8.4}{\rmdefault}{\mddefault}{\updefault}10384}}}}
\put(3322,-1754){\makebox(0,0)[lb]{\smash{{\SetFigFont{7}{8.4}{\rmdefault}{\mddefault}{\updefault}12007}}}}
\put(2809,-773){\makebox(0,0)[lb]{\smash{{\SetFigFont{7}{8.4}{\rmdefault}{\mddefault}{\updefault}7168}}}}
\put(3453,-918){\makebox(0,0)[lb]{\smash{{\SetFigFont{7}{8.4}{\rmdefault}{\mddefault}{\updefault}10216}}}}
\put(3474,-1087){\makebox(0,0)[lb]{\smash{{\SetFigFont{7}{8.4}{\rmdefault}{\mddefault}{\updefault}10282}}}}
\put(3207,-266){\makebox(0,0)[lb]{\smash{{\SetFigFont{7}{8.4}{\rmdefault}{\mddefault}{\updefault}3144}}}}
\put(1801,-1936){\makebox(0,0)[lb]{\smash{{\SetFigFont{7}{8.4}{\rmdefault}{\mddefault}{\updefault}Frequency (MHz)}}}}
\put(601,-4861){\makebox(0,0)[lb]{\smash{{\SetFigFont{7}{8.4}{\rmdefault}{\mddefault}{\updefault}Design 2}}}}
\put(601,-5011){\makebox(0,0)[lb]{\smash{{\SetFigFont{7}{8.4}{\rmdefault}{\mddefault}{\updefault}Design 4}}}}
\put(601,-5161){\makebox(0,0)[lb]{\smash{{\SetFigFont{7}{8.4}{\rmdefault}{\mddefault}{\updefault}Design 6}}}}
\put(1951,-4636){\makebox(0,0)[lb]{\smash{{\SetFigFont{7}{8.4}{\rmdefault}{\mddefault}{\updefault}$\text{area}\times\text{time}$ ($\text{slices}\cdot\mu\mathrm{s}$)}}}}
\put(601,-5536){\makebox(0,0)[lb]{\smash{{\SetFigFont{7}{8.4}{\rmdefault}{\mddefault}{\updefault}Design 2}}}}
\put(601,-5686){\makebox(0,0)[lb]{\smash{{\SetFigFont{7}{8.4}{\rmdefault}{\mddefault}{\updefault}Design 4}}}}
\put(601,-5836){\makebox(0,0)[lb]{\smash{{\SetFigFont{7}{8.4}{\rmdefault}{\mddefault}{\updefault}Design 6}}}}
\put(1951,-5311){\makebox(0,0)[lb]{\smash{{\SetFigFont{7}{8.4}{\rmdefault}{\mddefault}{\updefault}$\text{area}\times\text{time}^2$ ($\text{slices}\cdot\mu\mathrm{s}^2$)}}}}
\put(601,-2161){\makebox(0,0)[lb]{\smash{{\SetFigFont{7}{8.4}{\rmdefault}{\mddefault}{\updefault}Design 2}}}}
\put(601,-2311){\makebox(0,0)[lb]{\smash{{\SetFigFont{7}{8.4}{\rmdefault}{\mddefault}{\updefault}Design 4}}}}
\put(601,-2461){\makebox(0,0)[lb]{\smash{{\SetFigFont{7}{8.4}{\rmdefault}{\mddefault}{\updefault}Design 6}}}}
\put(601,-2836){\makebox(0,0)[lb]{\smash{{\SetFigFont{7}{8.4}{\rmdefault}{\mddefault}{\updefault}Design 2}}}}
\put(601,-2986){\makebox(0,0)[lb]{\smash{{\SetFigFont{7}{8.4}{\rmdefault}{\mddefault}{\updefault}Design 4}}}}
\put(601,-3136){\makebox(0,0)[lb]{\smash{{\SetFigFont{7}{8.4}{\rmdefault}{\mddefault}{\updefault}Design 6}}}}
\put(1801,-2611){\makebox(0,0)[lb]{\smash{{\SetFigFont{7}{8.4}{\rmdefault}{\mddefault}{\updefault}Quies. power (W)}}}}
\put(601,-4186){\makebox(0,0)[lb]{\smash{{\SetFigFont{7}{8.4}{\rmdefault}{\mddefault}{\updefault}Design 2}}}}
\put(601,-4336){\makebox(0,0)[lb]{\smash{{\SetFigFont{7}{8.4}{\rmdefault}{\mddefault}{\updefault}Design 4}}}}
\put(601,-4486){\makebox(0,0)[lb]{\smash{{\SetFigFont{7}{8.4}{\rmdefault}{\mddefault}{\updefault}Design 6}}}}
\put(1951,-3961){\makebox(0,0)[lb]{\smash{{\SetFigFont{7}{8.4}{\rmdefault}{\mddefault}{\updefault}Total power (W)}}}}
\put(601,-3511){\makebox(0,0)[lb]{\smash{{\SetFigFont{7}{8.4}{\rmdefault}{\mddefault}{\updefault}Design 2}}}}
\put(601,-3661){\makebox(0,0)[lb]{\smash{{\SetFigFont{7}{8.4}{\rmdefault}{\mddefault}{\updefault}Design 4}}}}
\put(601,-3811){\makebox(0,0)[lb]{\smash{{\SetFigFont{7}{8.4}{\rmdefault}{\mddefault}{\updefault}Design 6}}}}
\put(1951,-3286){\makebox(0,0)[lb]{\smash{{\SetFigFont{7}{8.4}{\rmdefault}{\mddefault}{\updefault}Dyn. power (W)}}}}
\put(2126,-2127){\makebox(0,0)[lb]{\smash{{\SetFigFont{7}{8.4}{\rmdefault}{\mddefault}{\updefault}123.12}}}}
\put(3433,-2282){\makebox(0,0)[lb]{\smash{{\SetFigFont{7}{8.4}{\rmdefault}{\mddefault}{\updefault}300.391}}}}
\put(3486,-2431){\makebox(0,0)[lb]{\smash{{\SetFigFont{7}{8.4}{\rmdefault}{\mddefault}{\updefault}307.787}}}}
\put(3455,-2795){\makebox(0,0)[lb]{\smash{{\SetFigFont{7}{8.4}{\rmdefault}{\mddefault}{\updefault}2.742}}}}
\put(3487,-2959){\makebox(0,0)[lb]{\smash{{\SetFigFont{7}{8.4}{\rmdefault}{\mddefault}{\updefault}2.786}}}}
\put(3480,-3123){\makebox(0,0)[lb]{\smash{{\SetFigFont{7}{8.4}{\rmdefault}{\mddefault}{\updefault}2.747}}}}
\put(2521,-3472){\makebox(0,0)[lb]{\smash{{\SetFigFont{7}{8.4}{\rmdefault}{\mddefault}{\updefault}0.957}}}}
\put(3476,-3631){\makebox(0,0)[lb]{\smash{{\SetFigFont{7}{8.4}{\rmdefault}{\mddefault}{\updefault}1.687}}}}
\put(2706,-3794){\makebox(0,0)[lb]{\smash{{\SetFigFont{7}{8.4}{\rmdefault}{\mddefault}{\updefault}1.123}}}}
\put(3120,-4141){\makebox(0,0)[lb]{\smash{{\SetFigFont{7}{8.4}{\rmdefault}{\mddefault}{\updefault}3.699}}}}
\put(3502,-4305){\makebox(0,0)[lb]{\smash{{\SetFigFont{7}{8.4}{\rmdefault}{\mddefault}{\updefault}4.453}}}}
\put(3200,-4461){\makebox(0,0)[lb]{\smash{{\SetFigFont{7}{8.4}{\rmdefault}{\mddefault}{\updefault}3.870}}}}
\put(3494,-4834){\makebox(0,0)[lb]{\smash{{\SetFigFont{7}{8.4}{\rmdefault}{\mddefault}{\updefault}29.38}}}}
\put(2055,-4989){\makebox(0,0)[lb]{\smash{{\SetFigFont{7}{8.4}{\rmdefault}{\mddefault}{\updefault}10.446}}}}
\put(2082,-5136){\makebox(0,0)[lb]{\smash{{\SetFigFont{7}{8.4}{\rmdefault}{\mddefault}{\updefault}11.19}}}}
\put(3496,-5507){\makebox(0,0)[lb]{\smash{{\SetFigFont{7}{8.4}{\rmdefault}{\mddefault}{\updefault}0.239}}}}
\put(1564,-5670){\makebox(0,0)[lb]{\smash{{\SetFigFont{7}{8.4}{\rmdefault}{\mddefault}{\updefault}0.034}}}}
\put(1590,-5813){\makebox(0,0)[lb]{\smash{{\SetFigFont{7}{8.4}{\rmdefault}{\mddefault}{\updefault}0.036}}}}
\end{picture}%